\documentclass[journal]{IEEEtran}
\usepackage{amsmath,amsfonts}
\usepackage{algorithmic}
\usepackage{algorithm}
\usepackage{array}
\usepackage{textcomp}
\usepackage{stfloats}
\usepackage{url}
\usepackage{verbatim}
\usepackage{graphicx}
\usepackage{cite}
\usepackage{xcolor}
\usepackage{multirow}
\usepackage{makecell}
\usepackage{amsthm}
\usepackage{subfigure}

\newtheorem{theorem}{\bf Theorem}

\newtheorem{lemma}{\bf Lemma}

\newcommand{\tabincell}[2]{\begin{tabular}{@{}#1@{}}#2\end{tabular}}

\begin{document}

\title{Adaptive Information Bottleneck Guided Joint Source and Channel Coding for Image Transmission\vspace{0.3cm}}

\author{Lunan Sun, Yang Yang, \emph{Member, IEEE}, Mingzhe Chen, \emph{Member, IEEE}, Caili Guo, \emph{Senior Member, IEEE},\\
Walid Saad, \emph{Fellow, IEEE} and H. Vincent Poor, \emph{Life Fellow, IEEE}
\thanks{
Manuscript received September 1, 2022; revised March 31, 2023. This work was supported in part by Fundamental Research Funds for the Central Universities under Grant 2021XD-A01-1; in part by the Key Program of National Natural Science Foundation of China under Grant 92067202; in part by Beijing Natural Science Foundation under Grant L222043; and in part by the U.S National Science Foundation under Grant CNS-2128448. (\emph{Corresponding author: Caili Guo.})

L. Sun and Y. Yang are with the Beijing
Key Laboratory of Network System Architecture and Convergence, School of Information and Communication Engineering, Beijing University of Posts and Telecommunications, Beijing 100876, China (e-mail: sunlunan@bupt.edu.cn; young0607@bupt.edu.cn).

M. Chen is with the Department of Electrical and Computer Engineering and Institute for Data Science and Computing, University of Miami, Coral Gables, FL, 33146 USA (e-mail: mingzhe.chen@miami.edu).

C. Guo is with the Beijing Laboratory of Advanced Information Net works, School of Information and Communication Engineering, Beijing University
of Posts and Telecommunications, Beijing 100876, China (e-mail: guocaili@bupt.edu.cn).

W. Saad is with the Wireless@VT Group, Bradley Department of Electrical and Computer Engineering, Virginia Tech, Arlington, VA, USA (email: walids@vt.edu).

H. V. Poor is with the Department of Electrical and Computer
Engineering, Princeton University, Princeton, NJ 08544 USA (e-mail: poor@princeton.edu)}
}


\maketitle

\begin{abstract}
Joint source and channel coding (JSCC) for image transmission has attracted increasing attention due to its robustness and high efficiency. However, the existing deep JSCC research mainly focuses on minimizing the distortion between the transmitted and received information under a fixed number of available channels. Therefore, the transmitted rate may be far more than its required minimum value. In this paper, an adaptive information bottleneck (IB) guided joint source and channel coding (AIB-JSCC) method is proposed for image transmission. The goal of AIB-JSCC is to reduce the transmission rate while improving the image reconstruction quality. In particular, a new IB objective for image transmission is proposed so as to minimize the distortion and the transmission rate. A mathematically tractable lower bound on the proposed objective is derived, and then, adopted as the loss function of AIB-JSCC. To trade off compression and reconstruction quality, an adaptive algorithm is proposed to adjust the hyperparameter of the proposed loss function dynamically according to the distortion during the training. Experimental results show that AIB-JSCC can significantly reduce the required amount of transmitted data and improve the reconstruction quality and downstream task accuracy.
\end{abstract}


\section{Introduction}
Shannon's information theory has laid the foundations of modern communication systems. In particular, according to Shannon's information theory, separate source and channel coding (SSCC) is optimal for a memoryless source and channel when the latency, complexity, and code length are not constrained\cite{shannon1948mathematical}. However, SSCC has several practical limitations. First, the theory is based on the assumption of potentially infinite code lengths, which are impossible in practice, and SSCC is suboptimal for finite code lengths. Also, to achieve theoretically optimal performance, maximum likelihood detection methods must be used, which can be, in general, NP-hard\cite{berlekamp1978inherent}, thus introducing very high computational complexity and leading to unacceptable latency. Furthermore, the envisioned sixth generation (6G) of wireless networks are expected to connect trillion-level devices and require 10 to 1000 times higher rates\cite{saad2019vision}. In addition, it is thought that 6G will support a wide range of services and applications\cite{9782523}, such as ugmented reality, medical imaging and autonomous vehicles\cite{chafii2022ten,9583918}, which have strict latency requirement\cite{9700645,9200631,wan2022road,chen2021distributed}. Therefore, SSCC may not be able to meet the requirements of 6G.

To address the above-mentioned challenges, joint source and channel coding (JSCC) has attracted increasing attention as a means to achieve reliable data transmission. Existing studies of JSCC can be classified into two types: traditional research based on mathematical models\cite{gallager1968information,gastpar2003code,cheung2000bit,heinen2005transactions} and deep learning (DL)-based research\cite{bourtsoulatze2019deep,kurka2020deepjscc,xu2021wireless,choi2019neural,song2020infomax}. Traditional JSCC research mainly relies on traditional source coding and channel coding theory while focusing on performance analysis under ideal channel or source assumptions\cite{gallager1968information,gastpar2003code}. Coding schemes, such as bit allocation algorithm\cite{cheung2000bit}, robust nonlinear block coding\cite{heinen2005transactions} have also been studied. However, these hand-crafted coding schemes may require additional tuning. Motivated by the impressive performance of DL in many domains such as computer vision\cite{Szegedy_2015_CVPR}, image compression\cite{NEURIPS2018_53edebc5}, and natural language processing\cite{devlin2018bert}, DL-based JSCC has been extensively studied\cite{bourtsoulatze2019deep,kurka2020deepjscc,xu2021wireless,choi2019neural,song2020infomax}, which can potentially support future semantic
communications\cite{9791409,wang2022performance}. Specifically, since images have larger dimensions than speech and text data, there is more information redundancy in images, and transmitting image data requires higher rate than transmitting speech and text data. Therefore, it is more challenging to design a DL-based JSCC system for image transmission.

\subsection{Related Works and Challenges}
The existing works on DL-based JSCC for image transmission model the communication system as a deep neural network (DNN)-based autoencoder\cite{bourtsoulatze2019deep,kurka2020deepjscc,xu2021wireless,choi2019neural,song2020infomax}. The main goal is jointly training the encoder and decoder to preserve information and improve the reconstruction quality. Minimizing the mean-squared error (MSE) between the input images and output reconstructions\cite{bourtsoulatze2019deep,kurka2020deepjscc,xu2021wireless} is commonly used to achieve this goal. In particular, the authors in \cite{bourtsoulatze2019deep} proposed an autoencoder-based JSCC architecture called deep JSCC that minimizes the MSE between the original images and the recovered images. Deep JSCC outperforms SSCC that combines JPEG or JPEG2000 with capacity-achieving channel codes. In \cite{kurka2020deepjscc}, the authors incorporated the channel output feedback into the transmission system and further improved the reconstruction quality of Deep JSCC. To address the variations of signal-to-noise ratios (SNRs) during transmission, the work in \cite{xu2021wireless} designed a novel JSCC scheme, which uses a channel-wise soft attention network to adapt automatically to various channel conditions. These existing works \cite{bourtsoulatze2019deep,kurka2020deepjscc,xu2021wireless} that use MSE as the distortion function to recover each pixel equally in image transmission may lose the information of important pixels thus reducing image reconstruction quality. In contrast, mutual information measures the distortion in terms of the distribution of images, which can emphasize key pixels and has stronger generalization ability. In \cite{choi2019neural}, a discrete variational autoencoder model is designed to maximize the mutual information between the source and noisy codewords. The authors in \cite{song2020infomax} developed a JSCC model to maximize the mutual information between the codewords and input image. Overall, the existing works on DL-based JSCC for image transmission aim at minimizing the distortion between the transmitted and received images by utilizing various distortion metrics such as MSE and mutual information as loss functions under a fixed number of achievable channels. While the works in \cite{bourtsoulatze2019deep,kurka2020deepjscc,xu2021wireless,choi2019neural,song2020infomax} are interesting, the theoretical minimum description length (or transmission rate) of the codewords to express source is neglected in the loss function. Therefore, the transmission rate may be much larger than the minimum required rate. A new form of loss function for JSCC, that simultaneously minimizes the transmission rate and the distortion deserves investigation.

Recently, the authors in \cite{tishby2000information} proposed an information-theoretic principle, termed information bottleneck (IB) to compress information and improve data fitting performance simultaneously by using mutual information between the codewords and the labels of the inputs as distortion. IB principle has been extensively applied in many domains including improving the performance of generalization and robustness\cite{alemi2016deep}, suppressing irrelevant features\cite{belinkov2020variational}, and dealing with domain shift\cite{du2020learning}. Since IB inherits the properties of RD theory, it can characterize the maximal compression ratio and the optimal features in theory\cite{tishby2015deep,lee2021reducing}. Therefore, we propose a novel IB-guided JSCC that can reduce the transmission rate for a given reconstruction quality. Here, we need to note that it is challenging to apply the IB principle in JSCC for image transmission since standard IB is particularly designed for supervised tasks, while a JSCC-based image transmission system can be viewed as an unsupervised data reconstruction task. Meanwhile, in an image transmission JSCC system, the distribution of the input images is usually unknown, and the dimension of the extracted codewords is large. Thus, the mutual information used in IB is intractable. Therefore, to apply the IB principle to JSCC for image transmission, two main challenges must be addressed:
\begin{itemize}
\item {How to design a proper form of IB for an image transmission JSCC system, which is unsupervised.}
\item{How to calculate the mutual information used in IB and obtain a tractable and differentiable IB objective.}
\end{itemize}

\subsection{Contributions}
The main contribution of this paper is an adaptive IB-guided JSCC (AIB-JSCC) scheme for image transmission to address the above issues. The major contributions of the paper can be summarized as follows:

\begin{itemize}
\item {We design a new form of IB objective that aims at simultaneously maximizing the mutual information between the received noisy codewords and the input images, and minimizing the mutual information between the transmitted codewords and the input images. Thus, the new IB objective enables the image transmission JSCC system to reduce the transmission rate while guaranteeing the reconstruction quality. To the best of the authors' knowledge, this is the first work that applies the IB principle to image transmission JSCC and provides a theoretically maximal compression ratio guidance for neural networks.}
\item{As the mutual information in the proposed IB objective is intractable for DNNs with high-dimensional features, we develop a new mathematically tractable and differentiable lower bound on the proposed IB objective via a variational lower bound and contrastive log-ratio upper bound (CLUB) on mutual information. The derived lower bound is used as the loss function of AIB-JSCC.}
\item {We propose an adaptive algorithm, which can adjust the hyperparameter of the proposed IB objective to balance the reconstruction distortion and the required transmission rate. In particular,  we first develop an algorithm to adjust the hyperparameter value dynamically by exploiting reconstruction error. Then, we derive an upper bound on the hyperparameter, which can prevent excessive information discarding in the transmitted codewords.
}
\end{itemize}

We compare AIB-JSCC with traditional SSCC and state-of-the-art JSCC methods and quantify the performance gain via extensive experiments. Simulation results show that AIB-JSCC significantly reduces the required storage space and the amount of transmitted image data.

The rest of this paper is organized as follows. In Section II, the system model is described. The proposed IB objective is presented in Section III. The adaptive IB algorithm is introduced in Section IV. In Section V, we provide extensive experimental results to verify the effectiveness of AIB-JSCC. Finally, the conclusions are drawn in Section VI.

\section{System Model}
In this section, we first describe the studied JSCC system model for image transmission. Then, we discuss the motivation for our work as well as the IB principle.

\begin{table*}[t!]
\setlength\tabcolsep{5pt}
\renewcommand\arraystretch{1.6}
\centering
\caption{\label{tab:notations}List of NOTATION.}
\begin{tabular}{|c|p{5.5cm}|c|p{5.5cm}|}
\hline

\textbf{Notation} & \textbf{Definition} & \textbf{Notation} & \textbf{Definition} \\
\hline
$N$ & The size of the images & $M$ & The length of codewords\\
\hline
$B$ & The sample number in a batch & $P$ & The number of parallel channels \\
\hline
${\rm{MSE}}\left[ w \right] $ & The MSE between the inputs and the reconstructions at the $w$-th epoch & ${\eta _{\boldsymbol{\varepsilon} }}\left(  \cdot  \right)$ &  Th transition function of BSC with error probability $\boldsymbol{\varepsilon}$\\
\hline
$\boldsymbol{\varphi}$ &  The parameters of the encoder neural network & $\boldsymbol{\theta}$ & The parameters of the decoder neural network\\
\hline
$\boldsymbol{\varepsilon}$ & The error probability of the channel & ${\varepsilon _k}$ & The error probability of the $k$-th subchannel\\
\hline
$\boldsymbol{x}$ & The input images & ${\boldsymbol{x}^{\left( i \right)}}$ &  The $i$-th input image\\
\hline
$\boldsymbol{\hat x}$ & The recovered images & ${\boldsymbol{\hat x}}^{\left( i \right)}$ &  The $i$-th recovered image\\
\hline
$x_j^{\left( i \right)}$ & The $j$-th pixel in the $i$-th input image & $\hat x_j^{\left( i \right)}$ & The $j$-th pixel in the $i$-th recovered image\\
\hline
$\boldsymbol{y}$ & The codewords to be transmitted & ${\boldsymbol{y}^{\left( i \right)}}$ &  The codewords extracted from ${\boldsymbol{x}^{\left( i \right)}}$\\
\hline
$y_m$ & The $m$-th element in $\boldsymbol{y}$ & $y_m^{\left( i \right)}$ &  The $m$-th element in ${\boldsymbol{y}^{\left( i \right)}}$\\
\hline
$\boldsymbol{\hat y}$ & The noisy codewords received by decoder & ${\boldsymbol{\hat y}^{\left( i \right)}}$ & The noisy codewords extracted from ${\boldsymbol{x}^{\left( i \right)}}$\\
\hline
${\hat y}_m$ & The $m$-th element in $\boldsymbol{\hat y}$ & ${\hat y}_m^{\left( i \right)}$ & The $m$-th element in ${\boldsymbol{\hat y}^{\left( i \right)}}$\\
\hline
${\boldsymbol{y}_{{\rm{ch}}k}}$ & The subcodewords to be transmitted across the $k$-th subchannel & ${{\boldsymbol{\hat y}}_{{\rm{ch}}k}}$ & The noisy subcodewords received across the $k$-th subchannel\\
\hline
${f_{\boldsymbol{\varphi}} }\left(  \cdot  \right)$ &  The output of the encoder neural network & ${E_{\boldsymbol{\varphi}}}\left(  \cdot  \right)$ & The encoding process from $\boldsymbol{x}$ to $\boldsymbol{y}$\\
\hline
${g_{\boldsymbol{\theta}} }\left(  \cdot  \right)$ & The output of the decoder neural network & ${D_{\boldsymbol{\theta}} }\left(  \cdot  \right)$ & The decoding process from $\boldsymbol{\hat y}$ to $\boldsymbol{\hat x}$\\
\hline
${\beta}$ & The hyperparameter of the proposed IB objective & ${{\beta _{{\rm{adp}}}}}$ & The value of $\beta$ calculated by the adaptive IB algorithm\\
\hline
${{\beta _{\max }}}$ & The upper bound on $\beta$ & $\beta _{\min }$ & The minimum value of $\beta$\\
\hline
\end{tabular}
\end{table*}

\subsection{System Model}\label{System Model_A}
\begin{figure}[t]
\centering{
\includegraphics[width=1\columnwidth]{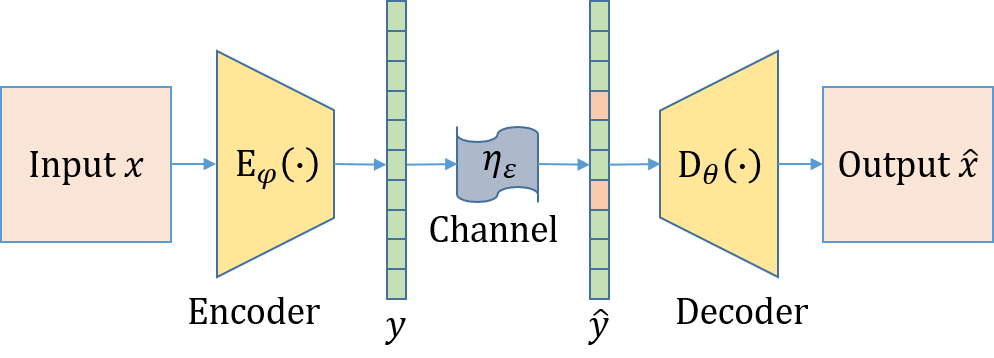}
}
\caption{\label{fig:illustration}An illustration of the JSCC system.}
\end{figure}

As shown in Fig. \ref{fig:illustration}, we consider a point-to-point image transmission system\cite{bourtsoulatze2019deep,kurka2020deepjscc,xu2021wireless,choi2019neural,song2020infomax}. An input image with size $H$ (Height) $\times$ $W$ (Width) $\times$ $C$ (Channel) is represented as a vector $\boldsymbol{x} \in {\mathbb{R}^{N}}$, where $\mathbb{R}$ represents the set of real numbers and $N = {H} \times {W} \times {C}$. The encoder encodes the image $\boldsymbol{x}$ into a binary codeword $\boldsymbol{y} \in {\left\{ {0,1} \right\}^M}$, where $M$ represents the length of the codeword $\boldsymbol{y}$ to be transmitted. The encoding function ${{\rm{E}}_{\boldsymbol{\varphi }}}:{\mathbb{R}^N} \to {\left\{ {0,1} \right\}^M}$ is parameterized by an encoder neural network with parameters $\boldsymbol{\varphi } $, and the encoding process can be expressed as
\begin{equation}
\label{model-0}
\boldsymbol{y} = {{\rm{E}}_{\boldsymbol{\varphi }} }\left( \boldsymbol{x} \right),
\end{equation}
where $\boldsymbol{y}$ is the JSCC codeword generated by encoding the source information and adding redundancy for error protection jointly. $\boldsymbol{y}$ is then transmitted across a noisy channel. To simplify the analysis, we do not consider concrete modulation, detection and decision schemes, and only consider the transmission of the codeword through a channel with a certain error probability, i.e., memoryless binary symmetric channel (BSC)\footnote{The BSC is a well-established and widely-used model in communication theory and information theory\cite{shannon1948mathematical,gallager1962low,cover1999elements} since it is a simple and tractable model for theoretical analysis.}. 
Here, we consider a BSC with error probability $0 \le {\boldsymbol{\varepsilon }} \le 0.5 $, denoted by ${\eta _{\boldsymbol{\varepsilon }} }:{\left\{ {0,1} \right\}^M} \to {\left\{ {0,1} \right\}^M}$. The channel output noisy codeword $\boldsymbol{\hat y} \in {\left\{ {0,1} \right\}^M}$ received by the decoder is expressed as
\begin{equation}
\label{model-1}
{\boldsymbol{\hat y}} = {\eta _{\boldsymbol{\varepsilon }} }\left( {\boldsymbol{y}} \right) = {\boldsymbol{y}} \oplus {\boldsymbol{z}},
\end{equation}
where $\boldsymbol{z} \sim {\rm{Bern}}\left({\boldsymbol{\varepsilon }} \right)$ represents the Bernoulli distributed noise of the considered channel, and $ \oplus $ represents modulo-2 addition\cite{romano2013minimum,podgorski1998estimation,huleihel2017quantize}. The channel capacity of the BSC with error probability ${\boldsymbol{\varepsilon }} $ is
\begin{equation}
\label{model-2}
{C_{{\rm{BSC}}}}\left( {\boldsymbol{\varepsilon }}  \right) = 1 - h\left( {\boldsymbol{\varepsilon }}  \right),
\end{equation}
where $h\left( {\boldsymbol{\varepsilon }}  \right) =- {\boldsymbol{\varepsilon }} \log {\boldsymbol{\varepsilon }}  - \left( {1 -{\boldsymbol{\varepsilon }} } \right)\log \left( {1 - {\boldsymbol{\varepsilon }} } \right)$ is the binary entropy function, and $\log \left( {\boldsymbol{x}} \right) \buildrel \Delta \over = {\log _2}\left( {\boldsymbol{x}} \right)$.

The decoder decodes the noisy codeword ${\boldsymbol{\hat y }}$ into reconstructed image $\boldsymbol{\hat x } \in {\mathbb{R}^{N}}$. The decoding function is parameterized by the decoder neural network parameters $\boldsymbol{\theta}$, and the decoding process is expressed as ${{\rm{D}}_{\boldsymbol{\theta}} }:{\left\{ {0,1} \right\}^M} \to {\mathbb{R}^{N}}$. The reconstructed image $\boldsymbol{\hat x }$ is
\begin{equation}
\label{model-3}
\boldsymbol{\hat x } = {{\rm{D}}_{\boldsymbol{\theta}} }\left( {\boldsymbol{\hat y}} \right) = {{\rm{D}}_{\boldsymbol{\theta}} }\left( {{\eta _{\boldsymbol{\varepsilon}} }\left( {{{\rm{E}}_{\boldsymbol{\varphi}} }\left( {\boldsymbol{x }} \right)} \right)} \right).
\end{equation}

The goal of the considered system is to determine the encoder and decoder parameters that minimize the average reconstruction error between $\boldsymbol{x}$ and $\boldsymbol{\hat x }$ while keeping the minimum description length (or transmission rate) of $\boldsymbol{y }$ to express $\boldsymbol{x }$ short.
\begin{figure*}[t!]
\centering{
\includegraphics[width=16cm]{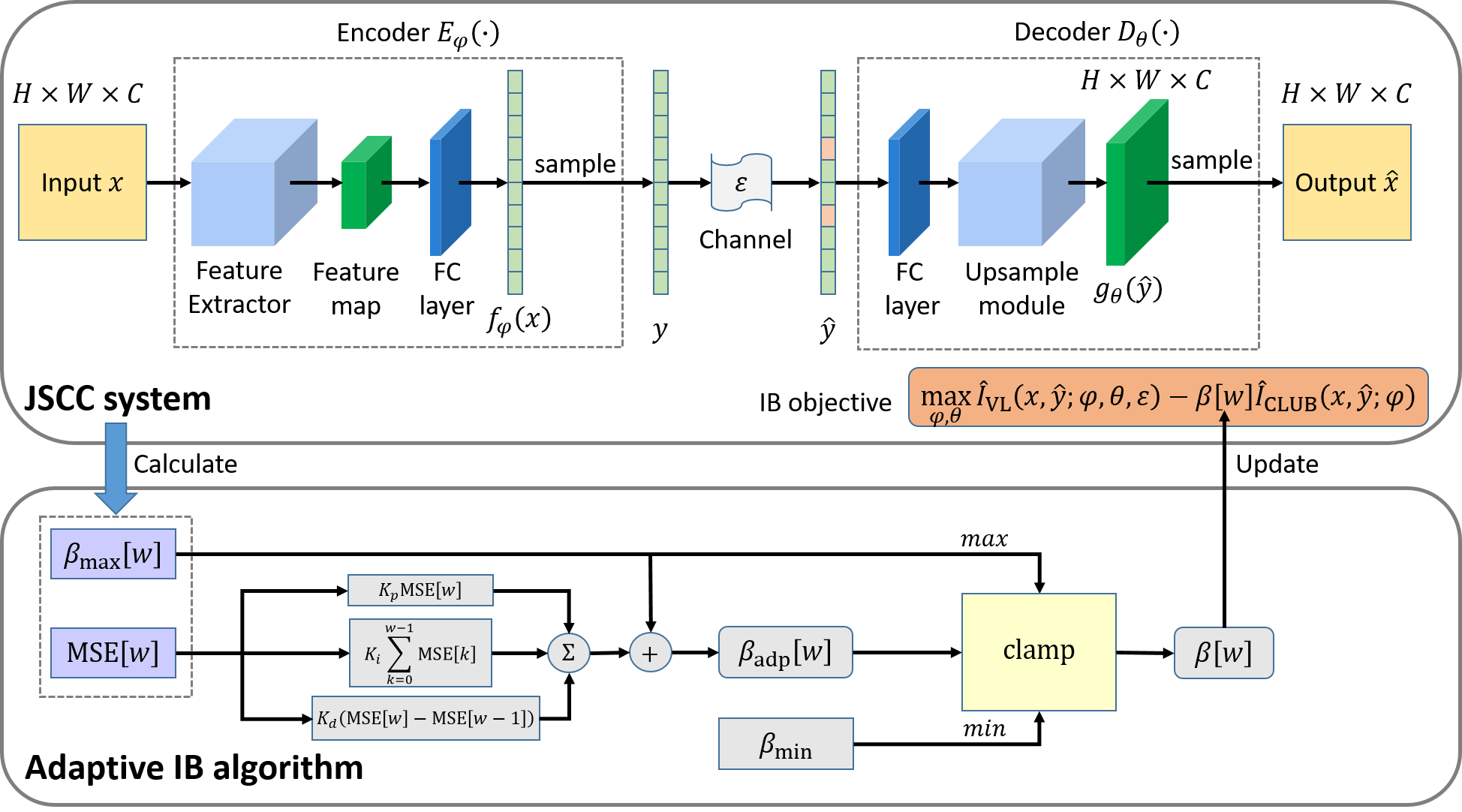}
}
\caption{\label{fig:AIB-JSCC}An illustration of our proposed AIB-JSCC. Top: JSCC system with the proposed IB objective. Bottom: adaptive IB algorithm. First, we train the JSCC system which consists of an autoencoder, by optimizing the proposed IB objective. Then, we adjust $\beta$ according to the proposed algorithm. Finally, we alternately change $\beta$ and train the network.}
\end{figure*}
\subsection{Motivation}\label{System Model_B}
Existing deep JSCC solutions for image transmission\cite{bourtsoulatze2019deep,kurka2020deepjscc,xu2021wireless,choi2019neural,song2020infomax} aim to minimize the distortion given a fixed number of available channels. However, they ignore the minimum description length (or transmission rate) of $\boldsymbol{y}$ to express $\boldsymbol{x}$, i.e., $I\left( {\boldsymbol{x};\boldsymbol{y}} \right)$ in the loss function. This motivates us to design a new loss function that can optimize both the distortion $d\left( {\boldsymbol{x},\boldsymbol{\hat x}} \right)$ and the transmission rate $I\left( {\boldsymbol{x};\boldsymbol{y}} \right)$ simultaneously. We resort to the IB principle. To extract the contained information of a target random variable $\boldsymbol{t}$ (e.g. label) in input $\boldsymbol{x}$, the authors in \cite{tishby2000information} used the mutual information between $\boldsymbol{y}$ and $\boldsymbol{t}$, ${I\left( {\boldsymbol{y};\boldsymbol{t}} \right)}$ as distortion measurement and proposed IB principle. The objective of IB is
\begin{equation}
\label{IB-2}
\mathop {\max }\limits_{p\left( {\boldsymbol{y}|\boldsymbol{x}} \right)} \left[ {{I\left( {\boldsymbol{y};\boldsymbol{t}} \right)} - \beta I\left( {\boldsymbol{x};\boldsymbol{y}} \right)} \right].
\end{equation}
The first term $I\left( {\boldsymbol{y};\boldsymbol{t}} \right)$ in (\ref{IB-2}) encourages $\boldsymbol{y}$ to predict $\boldsymbol{t}$, and the second term $I\left( {\boldsymbol{x};\boldsymbol{y}} \right)$ in (\ref{IB-2}) encourages $\boldsymbol{y}$ to compress the information related to $\boldsymbol{x}$. According to (\ref{IB-2}), the system can obtain the optimal $\boldsymbol{y}$ that is maximally compressed with a certain distortion\cite{lee2021reducing}. (\ref{IB-2}) is typically used as the loss function of supervised artificial intelligence tasks\cite{alemi2016deep,belinkov2020variational}, where $\boldsymbol{x}$ is the input image, $\boldsymbol{y}$ is the codeword, and $\boldsymbol{t}$ is the label of $\boldsymbol{x}$.

Even though the IB principle provides a new form of mutual information distortion, and can be used to guide the generation of the optimal features, the IB form in (\ref{IB-2}) is designed for supervised learning, and a label variable $\boldsymbol{t}$ is needed. Therefore, (\ref{IB-2}) cannot be applied to JSCC for image transmission directly, since image transmission is an unsupervised task.  Moreover, the value of hyperparameter $\beta$ in (\ref{IB-2}) needs to be carefully designed to balance prediction and compression. To solve these problems, we propose a new form of the IB objective that can minimize both the distortion and transmission rate for image transmission JSCC. We derive a tractable and differentiable lower bound on the proposed objective and use the bound as the loss function of AIB-JSCC for image transmission. An adaptive algorithm is also designed to dynamically adjust the hyperparameter $\beta$, so as to 
balance the compression and reconstruction quality.

\section{Proposed IB Objective for JSCC System}

This section first introduces the proposed IB objective for image transmission JSCC system. To obtain a tractable and differentiable form of the proposed IB objective, we then derive the lower bound of the proposed IB objective according to the variational lower bound and the upper bound of the mutual information.

\subsection{Proposed IB Objective}
The considered JSCC system is shown in Fig. \ref{fig:AIB-JSCC}, and it mainly consists of an encoder ${{\rm{E}}_{\boldsymbol{\varphi}} }\left(  \cdot  \right)$ block, a decoder block ${{\rm{D}}_{\boldsymbol{\theta}} }\left(  \cdot  \right)$, a channel block, and an adaptive IB algorithm block. In the considered system, IB principle is adopted to guide JSCC to achieve theoretically minimal transmission rate with a certain tasks distortion. However, the standard form of the IB principle shown as (\ref{IB-2}) is not applicable to the considered system, since image transmission is an unsupervised task without label. To overcome this problem, we propose a new form of the IB objective for image transmission JSCC as follows:
\begin{equation}
\label{methods-5}
\mathop {\max }\limits_{\boldsymbol{\varphi} ,\boldsymbol{\theta} } \left[ {I\left( {\boldsymbol{x};\boldsymbol{\hat y}} \right) - \beta I\left( {\boldsymbol{x};\boldsymbol{y}} \right)} \right].
\end{equation}

In the proposed IB objective, we use $I\left( {{\boldsymbol{x}};{\boldsymbol{\hat y}}} \right)$ to capture the reconstruction distortion between $\boldsymbol{x}$ and $\boldsymbol{\hat x}$. By maximizing $I\left( {{\boldsymbol{x}};{\boldsymbol{\hat y}}} \right)$, we can ensure that ${\boldsymbol{\hat y}}$ can capture the most useful information from ${\boldsymbol{x}}$. Hence, we can maximize $I\left( {{\boldsymbol{x}};{\boldsymbol{\hat x}}} \right)$, which is intractable due to unknown conditional probability $p\left( {{\boldsymbol{x}}|{\boldsymbol{\hat x}}} \right)$, via maximizing $I\left( {{\boldsymbol{x}};{\boldsymbol{\hat y}}} \right)$ since ${\boldsymbol{\hat x}}$ is reconstructed from ${\boldsymbol{\hat y}}$. However, since $I\left( {\boldsymbol{x};\boldsymbol{\hat y}} \right) \le I\left( {\boldsymbol{x};\boldsymbol{y}} \right)$, solely maximizing $I\left( {\boldsymbol{x};\boldsymbol{\hat y}} \right)$ may
result in severe information redundancy in $\boldsymbol{y}$, which implies that the system requires much higher transmission rate to transmit $\boldsymbol{y}$. Thus, we use the second term, which is the transmission rate over the channel, to compress the information in $\boldsymbol{y}$. We minimize $I\left( {{\boldsymbol{x}};{\boldsymbol{y}}} \right)$ so that the minimum description length (or transmission rate) of $\boldsymbol{y}$ to express $\boldsymbol{x}$ can be reduced. Although the sizes of $\boldsymbol{x}$ and $\boldsymbol{y}$ are fixed, the probability distribution of $\boldsymbol{y}$ can be optimized to minimize the minimum description length (or transmission rate) of $\boldsymbol{y}$ that is used to represent $\boldsymbol{x}$. We treat the proposed loss function as a joint optimization problem that integrates both reconstruction distortion minimization and transmission rate minimization. Utilizing (\ref{methods-5}) as the loss function, the JSCC system can accurately transmit images while minimizing the required transmission rate.

However, (\ref{methods-5}) still cannot be applied to the JSCC systems, since the mutual information terms $I\left( {\boldsymbol{x};\boldsymbol{\hat y}} \right)$ and $I\left( {\boldsymbol{x};\boldsymbol{y}} \right)$ in (\ref{methods-5}) are mathematically intractable due to the unknown $p\left( {\boldsymbol{x},\boldsymbol{y}} \right)$, $p\left( {\boldsymbol{x},\boldsymbol{\hat y}} \right)$, $p\left( \boldsymbol{x} \right)$, $p\left( \boldsymbol{
y} \right)$ and $p\left( \boldsymbol{\hat y} \right)$. To circumvent this challenge, we next derive the variational lower bound on $I\left( {\boldsymbol{x};\boldsymbol{\hat y}} \right)$ and estimate the upper bound on $I\left( {\boldsymbol{x};\boldsymbol{y}} \right)$.

\begin{figure}[t]
\centering{
\includegraphics[width=0.9\columnwidth]{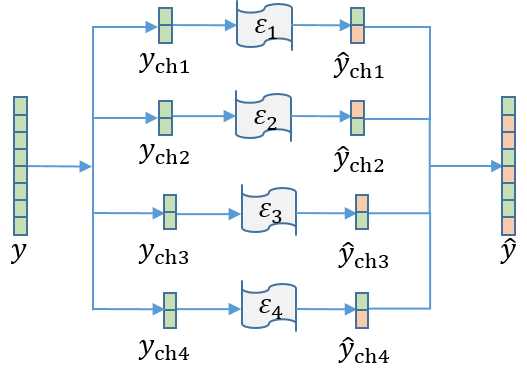}
}
\caption{\label{fig:parallel-channel}An illustration of the parallel-channel case with 4 subchannels. The codeword $\boldsymbol{y}$ is first equally divided into $4$ subcodewords, ${\boldsymbol{y}_{{\rm{ch1}}}}$, ${\boldsymbol{y}_{{\rm{ch2}}}}$, ${\boldsymbol{y}_{{\rm{ch3}}}}$ and ${\boldsymbol{y}_{{\rm{ch4}}}}$. These subcodewords are transmitted through their corresponding subchannel. At the receiver, the noisy subcodewords ${\boldsymbol{\hat y}_{{\rm{ch1}}}}$, ${\boldsymbol{\hat y}_{{\rm{ch2}}}}$, ${\boldsymbol{\hat y}_{{\rm{ch3}}}}$ and ${\boldsymbol{\hat y}_{{\rm{ch4}}}}$ are concatenated in order to obtain the noisy codeword $\boldsymbol{\hat y}$.}
\end{figure}

\subsection{Variational Lower Bound on $I\left( {\boldsymbol{x};\boldsymbol{\hat y}} \right)$} 
Instead of maximizing the true value of $I\left( {\boldsymbol{x};\boldsymbol{\hat y}} \right)$, we maximize its lower bound. We utilize the variational lower bound on  $I\left( {\boldsymbol{x};\boldsymbol{\hat y}} \right)$, which is obtained by \cite{agakov2004algorithm}
\begin{equation}
\label{methods-8}
\begin{aligned}
I\left( {\boldsymbol{x};\boldsymbol{\hat y}} \right) = & \underbrace {H\left( \boldsymbol{x} \right)}_{{\rm{constant}}} + \underbrace {{\mathbb{E}_{p\left( {\boldsymbol{x},\boldsymbol{\hat y}} \right)}}\log \left[ {\frac{{p\left( { \boldsymbol{x}|\boldsymbol{\hat y}} \right)}}{{q\left( { \boldsymbol{x}|\boldsymbol{\hat y}} \right)}}} \right]}_{{D_{KL}}\left( {p\left( {\boldsymbol{x}|\boldsymbol{\hat y}} \right)||q\left( {\boldsymbol{x}|\boldsymbol{\hat y}} \right)} \right) \ge {\rm{0}}}\\
& + \underbrace {{\mathbb{E}_{\boldsymbol{x}\sim p\left( \boldsymbol{x} \right)}}{\mathbb{E}_{\boldsymbol{\hat y}\sim p\left( {\boldsymbol{\hat y}|\boldsymbol{x}} \right)}}\log \left[ {q\left( {\boldsymbol{x}|\boldsymbol{\hat y}} \right)} \right]}_{{I_{{\rm{VL}}}}\left( {\boldsymbol{x};\boldsymbol{\hat y}} \right)}.
\end{aligned}
\end{equation} 
In (\ref{methods-8}), ${q\left( {\boldsymbol{x}|\boldsymbol{\hat y}} \right)}$ is the variational approximation of the true posterior ${p\left( {\boldsymbol{x}|\boldsymbol{\hat y}} \right)}$. The first term, $H\left( \boldsymbol{x} \right)$ is the entropy of the input images, which is a constant and cannot be optimized by neural networks. The second term is the Kullback-Leibler (KL) divergence between ${p\left( {\boldsymbol{x}|\boldsymbol{\hat y}} \right)}$ and ${q\left( {\boldsymbol{x}|\boldsymbol{\hat y}} \right)}$, which is positive. Since $H\left( \boldsymbol{x} \right) \ge 0$ and ${D_{KL}}\left( {{p\left( {\boldsymbol{x}|\boldsymbol{\hat y}} \right)}||{q\left( {\boldsymbol{x}|\boldsymbol{\hat y}} \right)}} \right) \ge {\rm{0}}$, the variational lower bound on $I\left( {\boldsymbol{x};\boldsymbol{\hat y}} \right)$ is ${{I_{{\rm{VL}}}}\left( {\boldsymbol{x};\boldsymbol{\hat y}} \right)}$, as defined in (\ref{methods-8}). The approximation error will be smaller if ${q\left( {\boldsymbol{x}|\boldsymbol{\hat y}} \right)}$ is closer to ${p\left( {\boldsymbol{x}|\boldsymbol{\hat y}} \right)}$, 

Since the conditional probability of $\boldsymbol{\hat y}$ given $\boldsymbol{x}$ depends on $\boldsymbol{\varphi}$ and $\boldsymbol{\varepsilon}$, we represent it as ${p\left( {\boldsymbol{\hat y}|\boldsymbol{x};\boldsymbol{\varphi} ,\boldsymbol{\varepsilon} } \right)}$. Denote the conditional probability of $\boldsymbol{\hat x}$ given $\boldsymbol{\hat y}$ as ${{p_{\boldsymbol{\theta}} }\left( {\boldsymbol{\hat x}|\boldsymbol{\hat y}} \right)}$, which is parameterized by the decoder neural network. We use ${{p_{\boldsymbol{\theta}} }\left( {\boldsymbol{\hat x}|\boldsymbol{\hat y}} \right)}$ as the variational approximation of the true posterior ${p\left( {\boldsymbol{x}|\boldsymbol{\hat y}} \right)}$. Since the BSC is discrete, ${p\left( {\boldsymbol{\hat y}|\boldsymbol{x};\boldsymbol{\varphi} ,\boldsymbol{\varepsilon} } \right)}$ is non-differentiable for $\boldsymbol{\varphi}$. Therefore, we sample $K$ noisy codewords $\boldsymbol{\hat y}$ for each input image $\boldsymbol{x}$ and use variational inference for Monte Carlo objectives (VIMCO)\cite{mnih2016variational}  to estimate ${{I_{{\rm{VL}}}}\left( {\boldsymbol{x};\boldsymbol{\hat y}} \right)}$ with low-variance gradients. The estimation ${{\hat I}_{{\rm{VL}}}}\left( {\boldsymbol{x},\boldsymbol{\hat y};{\boldsymbol{\varphi}} ,{\boldsymbol{\theta}} ,{\boldsymbol{\varepsilon}} } \right)$ can be expressed as\cite{choi2019neural,song2020infomax}
\begin{equation}
\label{VIMCO}
\begin{split}
{{\hat I}_{{\rm{VL}}}}&\left( {{\boldsymbol{x}},\boldsymbol{\hat y};{\boldsymbol{\varphi}} ,{\boldsymbol{\theta}} ,{\boldsymbol{\varepsilon}} } \right) =\\
& {\mathbb{E}_{p\left( {\boldsymbol{x}} \right)}}{\mathbb{E}_{p\left( {{{\boldsymbol{\hat y}}^{\left( 1 \right):\left( K \right)}}|{\boldsymbol{x}};{\boldsymbol{\varphi}} ,{\boldsymbol{\varepsilon}} } \right)}}\left[ {\log \frac{1}{K}\sum\limits_{i = 1}^K {{p_{\boldsymbol{\theta}} }\left( {{\boldsymbol{\hat x}}|{{\boldsymbol{\hat y}}^{\left( i \right)}}} \right)} } \right],
\end{split}
\end{equation}
where ${{{\boldsymbol{\hat y}}^{\left( i \right)}}}$ represents the $i$-th sample among $K$ samples. 

To calculate (\ref{VIMCO}), we need to know $p\left( {\boldsymbol{\hat y}|{\boldsymbol{x}};{\boldsymbol{\varphi}} ,{\boldsymbol{\varepsilon}} } \right)$ and ${p_{\boldsymbol{\theta}} }\left( {{\boldsymbol{\hat x}}|\boldsymbol{\hat y}} \right)$ first. According to (\ref{model-0}), (\ref{model-1}), and (\ref{model-3}), a Markov chain, ${\boldsymbol{x}} \to {\boldsymbol{y}} \to \boldsymbol{\hat y} \to {\boldsymbol{\hat x}}$ exists in the JSCC system. Thus, the joint probability $p\left( {{\boldsymbol{x}},{\boldsymbol{\hat x}},{\boldsymbol{y}},\boldsymbol{\hat y}} \right)$ can be modelled as
\begin{equation}
\label{methods-1}
p\left( {{\boldsymbol{x}},{\boldsymbol{\hat x}},{\boldsymbol{y}},\boldsymbol{\hat y}} \right) = p\left( {\boldsymbol{x}} \right){p_{\boldsymbol{\varphi}} }\left( {{\boldsymbol{y}}|{\boldsymbol{x}}} \right){p_{\boldsymbol{\varepsilon}} }\left( {\boldsymbol{\hat y}|{\boldsymbol{y}}} \right){p_{\boldsymbol{\theta}} }\left( {{\boldsymbol{\hat x}}|\boldsymbol{\hat y}} \right),
\end{equation}
where ${p_{\boldsymbol{\varphi}} }\left( {{\boldsymbol{y}}|{\boldsymbol{x}}} \right)$ is the conditional probability of $\boldsymbol{y}$ given $\boldsymbol{x}$, which is parameterized by the encoder neural network, and ${p_{\boldsymbol{\varepsilon}} }\left( {\boldsymbol{\hat y}|{\boldsymbol{y}}} \right)$ is the channel transition probability for BSC. Here, we consider the following two types of BSC:
\begin{itemize}
\item {
Single-channel: When the system utilizes single-carrier modulation such as binary phase shift keying (BPSK), the error probability of different bands is the same. This scenario is referred to as a single-channel scenario.}
\item{Parallel-channel: When the system utilizes multicarrier modulation, e.g., orthogonal frequency division multiplexing (OFDM) to resist channel fading, the total available bandwidth is divided into non-overlapping bands, and the transmitted data stream will be divided into substreams and sent via parallel bands. In this case, each of the parallel bands has a different error probability. This scenario is referred to as a parallel-channel scenario. For instance, a parallel-channel scenario with $4$ subchannels is shown in Fig. \ref{fig:parallel-channel}}.
\end{itemize}

We assume that the system has $P$ parallel subchannels with equal bandwidth $\frac{M}{P}$, where $P \in {\mathbb{N}^ + }$ and $M$ is the length of $\boldsymbol{y}$. Note that $P = 1$ represents the single-channel case. Denote the error probabilities of different bands as ${\boldsymbol{\varepsilon}}  = \left\{ {{\varepsilon _1},{\varepsilon _2}, \ldots ,{\varepsilon _P}} \right\}$, the channel transition probability is
\begin{equation}
\label{JSAC-3}
{p_{\boldsymbol{\varepsilon}} }\left( {\boldsymbol{\hat y}|{\boldsymbol{y}}} \right) = \prod\limits_{k = 1}^P {\prod\limits_{m = \frac{M}{P}\left( {k - 1} \right) + 1}^{\frac{M}{P}k} {{\varepsilon _k}^{{y_m} \oplus {{\hat y}_m}}{{\left( {1 - {\varepsilon _k}} \right)}^{{y_m} \oplus {{\hat y}_m} \oplus 1}}} },
\end{equation}
where ${{\hat y}_m}$ represents the $m$-th element of $\boldsymbol{\hat y}$. Note that the proposed AIB-JSCC is also applicable to JSCC systems with arbitrary discrete memoryless channels (DMCs). 

Let ${{f_{\boldsymbol{\varphi}} }\left( {\boldsymbol{x}} \right)}$ denote the output of the encoder neural network when the input is $\boldsymbol{x}$.  Since ${\boldsymbol{y}}$ is a binary codeword, we use the Bernoulli distribution to parameterize ${p_{\boldsymbol{\varphi }}}({\boldsymbol{y}}|{\boldsymbol{x}})$. To reduce the redundancy between any two elements of ${\boldsymbol{y}}$, we assume that the elements in ${\boldsymbol{y}}$ are independent of each other, and ${{f_{\boldsymbol{\varphi}} }\left( \boldsymbol{x} \right)}$ is treated as the parameters of this Bernoulli distribution, i.e., ${\boldsymbol{y}} = {{\rm{E}}_{\boldsymbol{\varphi}} }\left( \boldsymbol{x} \right) \sim {\rm{Bern}}\left( {{f_{\boldsymbol{\varphi}} }\left( \boldsymbol{x} \right)} \right)$. Then, ${p_{\boldsymbol{\varphi}} }\left( {{\boldsymbol{y}}|\boldsymbol{x}} \right)$ is
\begin{equation}
\label{methods-2}
\begin{aligned}
{p_{\boldsymbol{\varphi}} }\left( {{\boldsymbol{y}}|\boldsymbol{x}} \right) & =  \prod\limits_{m = 1}^M {{p_{\boldsymbol{\varphi}} }\left( {{y_m}|\boldsymbol{x}} \right)} \\
& = \prod\limits_{m = 1}^M {{{\left( {{f_{\boldsymbol{\varphi}} }\left( \boldsymbol{x} \right)} \right)}^{{y_m}}}{{\left( {1 - {f_{\boldsymbol{\varphi}} }\left( \boldsymbol{x} \right)} \right)}^{1 - {y_m}}}},
\end{aligned}
\end{equation}
where ${{y_m}}$ represents the $m$-th element of $\boldsymbol{y}$. The channel state information (CSI) is assumed to be perfectly estimated. Hence, both the encoder and the decoder know the accurate $\boldsymbol{\varepsilon}$. We can compute $p\left( {\boldsymbol{\hat y}|\boldsymbol{x};{\boldsymbol{\varphi}} ,{\boldsymbol{\varepsilon}} } \right)$ by marginalizing over $\boldsymbol{y}$ as
\begin{equation}
\label{TPR}
p\left( {{\boldsymbol{\hat y}}|{\boldsymbol{x}};{\boldsymbol{\varphi}} ,{\boldsymbol{\varepsilon}} } \right) = \sum\limits_{\boldsymbol{y} \in {{\left\{ {0,1} \right\}}^M}} {{p_{\boldsymbol{\varphi}} }\left( {{\boldsymbol{y}|\boldsymbol{x}}} \right){p_{\boldsymbol{\varepsilon}} }\left( {{\boldsymbol{\hat y}}|\boldsymbol{y}} \right)}.
\end{equation}
Then, $p\left( {{\boldsymbol{\hat y}}|{\boldsymbol{x}};{\boldsymbol{\varphi}} ,{\boldsymbol{\varepsilon}} } \right)$ is formulated as:

\begin{equation}
\label{JSAC-4}
p\left( {{\boldsymbol{\hat y}}|{\boldsymbol{x}};{\boldsymbol{\varphi}} ,{\boldsymbol{\varepsilon}} } \right) = \prod\limits_{k = 1}^P {\prod\limits_{m = \frac{M}{P}\left( {k - 1} \right) + 1}^{\frac{M}{P}k} {{{\left( {{\xi _k}\left( {\boldsymbol{x}} \right)} \right)}^{{{\hat y}_m}}}{{\left( {1 - {\xi _k}\left( {\boldsymbol{x}} \right)} \right)}^{1 - {{\hat y}_m}}}} },
\end{equation}
where ${\xi _k}\left( {\boldsymbol{x}} \right) = {f_{\boldsymbol{\varphi}} }\left( {\boldsymbol{x}} \right) - 2{f_{\boldsymbol{\varphi}} }\left( {\boldsymbol{x}} \right){\varepsilon _k} + {\varepsilon _k}$.  From (\ref{JSAC-4}), we can observe that $p\left( {{\boldsymbol{\hat y}}|{\boldsymbol{x}};{\boldsymbol{\varphi}} ,{\boldsymbol{\varepsilon}} } \right)$ follows multivariate independent Bernoulli distribution with parameters ${\xi _k}\left( {\boldsymbol{x}} \right)$. 

Since ${\boldsymbol{x}}$ can be normalized to a real-value vector where each element value is  within $0$ and $1$, we use a Gaussian distribution to model ${p_{\boldsymbol{\theta }}}({\boldsymbol{\hat x}}|{\boldsymbol{\hat y}})$ such that ${p_{\boldsymbol{\theta }}}({\boldsymbol{\hat x}}|{\boldsymbol{\hat y}})$ is differential with respect to ${\boldsymbol{\theta }}$. Let ${g_{\boldsymbol{\theta}} }\left( {{\boldsymbol{\hat y}}} \right)$ represent the output of the decoder neural network when the input of the decoder is $\boldsymbol{\hat y}$. We assume that the average of ${p_{\boldsymbol{\theta }}}({\boldsymbol{\hat x}}|{\boldsymbol{\hat y}})$ is ${g_{\boldsymbol{\theta}} }\left( {{\boldsymbol{\hat y}}} \right)$\cite{choi2019neural,song2020infomax}, i.e., ${\boldsymbol{\hat x}} = {{\rm{D}}_{\boldsymbol{\theta}} }\left( {{\boldsymbol{\hat y}}} \right) \sim {\cal N}\left( {{g_{\boldsymbol{\theta}} }\left( {{\boldsymbol{\hat y}}} \right),\boldsymbol{I}} \right)$, where $\cal N$ represents the Gaussian distribution. Then, we have

\begin{equation}
\label{decoder-1}
{p_{\boldsymbol{\theta}} }\left( {{\boldsymbol{\hat x}}|{\boldsymbol{\hat y}}} \right) = \prod\limits_{i = 1}^N {\frac{1}{{\sqrt {2\pi } }}\exp \left( {\frac{{{{{{\hat x}}}_i} - {g_{\boldsymbol{\theta}} }{{\left( {{\boldsymbol{\hat y}}} \right)}_i}}}{2}} \right)},
\end{equation}
where ${{\hat x}_i}$ is the $i$-th pixel of $\boldsymbol{\hat x}$, and ${{g_{\boldsymbol{\theta}} }{{\left( {{\boldsymbol{\hat y}}} \right)}_i}}$ is the corresponding pixel in ${{g_{\boldsymbol{\theta}} }\left( \boldsymbol{{\boldsymbol{\hat y}}} \right)}$. Then ${{\hat I}_{{\rm{VL}}}}\left( {\boldsymbol{x},\boldsymbol{\hat y};\boldsymbol{\varphi} ,\boldsymbol{\theta} ,\boldsymbol{\varepsilon} } \right)$ can be calculated by introducing (\ref{JSAC-4}) and (\ref{decoder-1}) into (\ref{VIMCO}).

\subsection{Upper Bound on $I\left( {\boldsymbol{x};\boldsymbol{y}} \right)$} 
Next, we derive the applicable form of $I\left( {\boldsymbol{x};\boldsymbol{y}} \right)$ in the considered system. Since $I\left( {\boldsymbol{x};\boldsymbol{y}} \right)$ is mathematically intractable, we minimize its upper bound instead. However, since we do not constrain the distribution of $\boldsymbol{y}$, the popular variational upper bound (VUB)\cite{alemi2016deep}, $KL\left( {p\left( {{\boldsymbol{y}|\boldsymbol{x}}} \right)|r\left( {\boldsymbol{y}} \right)} \right)$, cannot be used, where ${r\left( {\boldsymbol{y}} \right)}$ is an approximation of $p\left( {\boldsymbol{y}} \right)$. Therefore, we exploit another upper bound on mutual information called CLUB as \cite{cheng2020club}
\begin{equation}
\label{methods-9}
\begin{aligned}
{I_{{\rm{CLUB}}}}\left( {{\boldsymbol{x}},{\boldsymbol{y}};{\boldsymbol{\varphi}} } \right) = &\ {\mathbb{E}_{p\left( {\boldsymbol{x},\boldsymbol{y}} \right)}}\left[ {\log {p_{\boldsymbol{\varphi}} }\left( {{\boldsymbol{y}|\boldsymbol{x}}} \right)} \right]\\
& - {\mathbb{E}_{p\left( {\boldsymbol{x}} \right)}}{\mathbb{E}_{p\left( {\boldsymbol{y}} \right)}}\left[ {\log {p_{\boldsymbol{\varphi}} }\left( {{\boldsymbol{y}|\boldsymbol{x}}} \right)} \right].
\end{aligned}
\end{equation}
Let $B$ denote the number of independent sample pairs $\left\{ {\left( {{\boldsymbol{x}^{\left( i \right)}},{\boldsymbol{y}^{\left( i \right)}}} \right)} \right\}_{i = 1}^B$, where ${{\boldsymbol{x}^{\left( i \right)}}}$ represents the $i$-th image, and ${{\boldsymbol{y}^{\left( i \right)}}}$ represents the corresponding $i$-th codeword. Then ${I_{{\rm{CLUB}}}}\left( {{\boldsymbol{x}},{\boldsymbol{y}};{\boldsymbol{\varphi}} } \right)$ can be estimated by the Monte Carlo method as:
\begin{equation}
\label{model-B-6}
\begin{aligned}
{{\hat I}_{{\rm{CLUB}}}}\left( {{\boldsymbol{x}},{\boldsymbol{y}};{\boldsymbol{\varphi}} } \right) = &\  \frac{1}{B}\sum\limits_{i = 1}^B {\log {p_{\boldsymbol{\varphi}} }\left( {{\boldsymbol{y}^{\left( j \right)}}|{\boldsymbol{x}^{\left( i \right)}}} \right)}\\
& - \frac{1}{{{B^2}}}\sum\limits_{i = 1}^B {\sum\limits_{j = 1}^B {\log {p_{\boldsymbol{\varphi}} }\left( {{\boldsymbol{y}^{\left( j \right)}}|{\boldsymbol{x}^{\left( i \right)}}} \right)} }.
\end{aligned}
\end{equation}
Since ${{p_{\boldsymbol{\varphi}} }\left( {{\boldsymbol{y}|\boldsymbol{x}}} \right)}$ is the probability of a Bernoulli distribution, ${{\hat I}_{{\rm{CLUB}}}}\left( {{\boldsymbol{x}},{\boldsymbol{y}};{\boldsymbol{\varphi}} } \right)$ is tractable and differentiable. Thus, instead of minimizing the true value of $I\left( {\boldsymbol{x};\boldsymbol{\hat y}} \right)$, we minimize ${{\hat I}_{{\rm{CLUB}}}}\left( {{\boldsymbol{x}},{\boldsymbol{y}};{\boldsymbol{\varphi}} } \right)$.

Overall, by replacing $I\left( {\boldsymbol{x};\boldsymbol{\hat y}} \right)$ and $I\left( {\boldsymbol{x};\boldsymbol{y}} \right)$ in (\ref{methods-5}) with ${{\hat I}_{{\rm{VL}}}}\left( {\boldsymbol{x},{\boldsymbol{\hat y}};{\boldsymbol{\varphi}},{\boldsymbol{\theta}},{\boldsymbol{\varepsilon}} } \right)$ in (\ref{VIMCO}) and ${{\hat I}_{{\rm{CLUB}}}}\left( {{\boldsymbol{x}},{\boldsymbol{y}};{\boldsymbol{\varphi}} } \right)$ in (\ref{model-B-6}), respectively, we can obtain a tractable and differential form of IB objective for the JSCC system as:
\begin{equation}
\label{loss}
\mathop {\max }\limits_{{\boldsymbol{\varphi}} ,{\boldsymbol{\theta}} } \left[ {{{\hat I}_{{\rm{VL}}}}\left( {x,{\boldsymbol{\hat y}};{\boldsymbol{\varphi}} ,{\boldsymbol{\theta}} ,{\boldsymbol{\varepsilon}} } \right) - \beta {{\hat I}_{{\rm{CLUB}}}}\left( {{\boldsymbol{x}},{\boldsymbol{y}};{\boldsymbol{\varphi}} } \right)} \right].
\end{equation}
Even though (\ref{loss}) can be used for training, the value of $\beta$ needs to be carefully optimized, which controls the trade-off between the compression level and the reconstruction quality. Therefore, in Section \ref{Adaptive IB Algorithm}, we further propose an adaptive IB algorithm to determine the appropriate value of $ \beta$.
\section{Adaptive IB Algorithm}\label{Adaptive IB Algorithm}
This section first proposes an adaptive IB algorithm to select appropriate value of $ \beta$ according to the distortion of reconstruction during the training process. We then describe the whole training process of AIB-JSCC which combines the proposed IB objective and the adaptive IB algorithm.
\subsection{Adaptive IB Algorithm}
Since the values of $I\left( {\boldsymbol{x};\boldsymbol{\hat y}} \right)$ and $I\left( {\boldsymbol{x};\boldsymbol{y}} \right)$ change during the training process, it is necessary to alter $\beta$ accordingly so as to balance $I\left( {\boldsymbol{x};\boldsymbol{\hat y}} \right)$ and $I\left( {\boldsymbol{x};\boldsymbol{y}} \right)$.  To adaptively determine the value of $\beta$ in each epoch, we propose a proportional-integral-differential (PID) control based algorithm, which determines the current value of $\beta$ by analyzing the past errors and predicting future errors. The discrete form of PID controller can be expressed as\cite{li2021learning}
\begin{equation}
\label{Changeable IB-1}
\beta \left[ w \right] = {K_p}e\left[ w \right] - {K_i}\sum\limits_{k = 0}^{w - 1} {e\left[ k \right]}  - {K_d}\left( {e\left[ w \right] - e\left[ {w - 1} \right]} \right),
\end{equation}
where $\beta \left[ w \right]$ is the output of the controller at time $w$. ${K_p}$, ${K_i}$, ${K_d}$ and error $e\left[ w \right]$ are the proportional gain, the integral gain, the differential gain and the difference between the actual value and the desired value at time $w$, respectively. In addition, ${K_p}e\left[ w \right]$ is the proportional (P) term, which responds to the change of error quickly and provides a global control proportional to the error; ${K_i}\sum\limits_{k = 0}^{w - 1}$ is the integral (I) term, which continues to increase as long as the error is greater than $0$ and is used to eliminate steady-state errors; ${K_d}\left( {e\left[ w \right] - e\left[ {w - 1} \right]} \right)$ is the differential (D) term, which can reduce the overshoot and improve the system's stability and transient response\cite{li2021learning}.  The PID controller continuously calculates error $e\left[ w \right]$ and the weighted sum of these three terms, and then applies a correction on the system to reduce the error $e\left[ w \right]$.  We employ (\ref{Changeable IB-1}) to adjust $\beta$ at the end of each epoch. 

However, before applying (\ref{Changeable IB-1}) to the AIB-JSCC system, an upper bound of $\beta$ must to derived. This is because if $ \beta $ is excessively large, ${I\left( {\boldsymbol{x};\boldsymbol{y}} \right)}$ will dominate the loss function, and $\boldsymbol{y}$ will aggressively discard information of $\boldsymbol{x}$, leading to the loss of useful information. Considering an extreme case when $ \beta$ approaches positive infinity, ${I\left( {\boldsymbol{x};\boldsymbol{y}} \right)}$ will approach 0, and in this case, the global optimal encoder distribution may be ${p_{\boldsymbol{\varphi}} }\left( {{\boldsymbol{y}|\boldsymbol{x}}} \right) = p\left( {\boldsymbol{y}} \right)$. That means $\boldsymbol{y}$ becomes independent of $\boldsymbol{x}$. In this case, $\boldsymbol{y}$ and $\boldsymbol{\hat y}$ contain no information about $\boldsymbol{x}$, i.e. $I\left( {\boldsymbol{x};\boldsymbol{\hat y}} \right) = I\left( {\boldsymbol{x};\boldsymbol{y}} \right) = 0$, and reconstructing $\boldsymbol{x}$ from $\boldsymbol{\hat y}$ becomes infeasible. Therefore, it is necessary to limit $ \beta $ below an upper bound before applying PID controller, as shown in the following lemma.
\begin{lemma}\label{le:1}\emph{The condition that ${p_{\boldsymbol{\varphi}} }\left( {{\boldsymbol{y}|\boldsymbol{x}}} \right) = p\left( {\boldsymbol{y}} \right)$ is not a local optimum for the IB objective is \cite{wu2020learnability}
\begin{equation}
\label{Changeable IB-3}
\beta < {\beta _{\max }} = \mathop {\sup }\limits_{\boldsymbol{x} \to \boldsymbol{y} \to {\boldsymbol{\hat y}}} \frac{{I\left( {\boldsymbol{x};\boldsymbol{\hat y}} \right)}}{{I\left( {\boldsymbol{x};\boldsymbol{y}} \right)}}.
\end{equation}}
\end{lemma}
\begin{proof} See Appendix.
\end{proof}
According to Lemma \ref{le:1}, we derive the estimated upper bound on $\beta$ of the proposed AIB-JSCC in the following theorem.
\begin{theorem}\label{th:1}\emph{
For $B$ pairs $\left\{ {\left( {{\boldsymbol{x}^{\left( i \right)}},{\boldsymbol{y}^{\left( i \right)}},{{{\boldsymbol{\hat y}}}^{\left( i \right)}},{{{\boldsymbol{\hat x}}}^{\left( i \right)}}} \right)} \right\}_{i = 1}^B$, the estimated upper bound on $\beta$ is
\begin{equation}
\label{theorem_betamax_1}
{\beta _{\max }} = \frac{{{{\hat I}_{\boldsymbol{x},{\boldsymbol{\hat y}}}}}}{{{{\hat I}_{{\boldsymbol{x}},{\boldsymbol{y}}}}}},
\end{equation}
where
\begin{equation}
\label{theorem_betamax_2}
\begin{aligned}
{{\hat I}_{{\boldsymbol{x}},{\boldsymbol{y}}}} = & \sum\limits_{m = 1}^M {H\left( {\frac{1}{B}\sum\limits_{i = 1}^B {p\left( {y_m^{\left( i \right)}|{\boldsymbol{x}^{\left( i \right)}}} \right)} } \right)}  \\
& - \frac{1}{B}\sum\limits_{m = 1}^M {\sum\limits_{i = 1}^B {H\left( {p\left( {y_m^{\left( i \right)}|{\boldsymbol{x}^{\left( i \right)}}} \right)} \right)} },
\end{aligned}
\end{equation}
and
\begin{equation}
\label{theorem_betamax_3}
\begin{aligned}
{{\hat I}_{\boldsymbol{x},{\boldsymbol{\hat y}}}} =& \sum\limits_{m = 1}^M {H\left( {\frac{1}{B}\sum\limits_{i = 1}^B {p\left( {\hat y_m^{\left( i \right)}|{\boldsymbol{x}^{\left( i \right)}}} \right)} } \right)}  \\
& - \frac{1}{B}\sum\limits_{m = 1}^M {\sum\limits_{i = 1}^B {H\left( {p\left( {\hat y_m^{\left( i \right)}|{\boldsymbol{x}^{\left( i \right)}}} \right)} \right)} }.    
\end{aligned}
\end{equation}}
\end{theorem}
\begin{proof}
At the end of each epoch, ${I\left( {\boldsymbol{x};\boldsymbol{\hat y}} \right)}$ and ${I\left( {\boldsymbol{x};\boldsymbol{y}} \right)}$ are fixed since neural networks of the encoder and decoder are fixed. Therefore, according to Lemma \ref{le:1}, ${\beta _{\max }} = \frac{{I\left( {\boldsymbol{x};\boldsymbol{\hat y}} \right)}}{{I\left( {\boldsymbol{x};\boldsymbol{y}} \right)}}$. We then estimate ${I\left( {\boldsymbol{x};\boldsymbol{\hat y}} \right)}$ and ${I\left( {\boldsymbol{x};\boldsymbol{y}} \right)}$ according to the definition of the mutual information, i.e., $I\left( {\boldsymbol{x};\boldsymbol{\hat y}} \right) = H\left( {{\boldsymbol{\hat y}}} \right) - H\left( {{\boldsymbol{\hat y}}|\boldsymbol{x}} \right)$, and $I\left( {{\boldsymbol{x}};{\boldsymbol{y}}} \right) = H\left( {\boldsymbol{y}} \right) - H\left( {{\boldsymbol{y}|\boldsymbol{x}}} \right) $. We estimate ${H\left( {\boldsymbol{y}} \right)}$ and ${H\left( {{\boldsymbol{y}|\boldsymbol{x}}} \right)}$ separately. To obtain ${H\left( {\boldsymbol{y}} \right)}$, we calculate the probability of the $m$-th element of $\boldsymbol{y}$, $p\left( {{y_m}} \right)$, by,
\begin{equation}
\label{bound-1}
p\left( {{y_m}} \right) = \frac{1}{B}\sum\limits_{i = 1}^B {p\left( {y_m^{\left( i \right)}|{\boldsymbol{x}^{\left( i \right)}}} \right)},
\end{equation}
where ${y_m^{\left( i \right)}}$ represents the $m$-th element of the codeword of the $i$-th input image ${{\boldsymbol{x}^{\left( i \right)}}}$. Since the elements in $\boldsymbol{y}$ are assumed to be independent, the entropy of $\boldsymbol{y}$ is equal to the sum of the entropies of all elements. Besides, we assume $p\left( {\boldsymbol{x}} \right) = \frac{1}{B}$, and we have
\begin{equation}
\label{bound-2}
\begin{aligned}
H\left( {\boldsymbol{y}} \right) & = \sum\limits_{m = 1}^M {H\left( {{y_m}} \right)}  \approx \hat H\left( {\boldsymbol{y}} \right)\\
& = \sum\limits_{m = 1}^M {H\left( {\frac{1}{B}\sum\limits_{i = 1}^B {p\left( {y_m^{\left( i \right)}|{\boldsymbol{x}^{\left( i \right)}}} \right)} } \right)}.
\end{aligned}
\end{equation}
Substituting (\ref{methods-2}) into (\ref{bound-2}), $H\left( {\boldsymbol{y}} \right)$ can be calculated. To calculate $H\left( {{\boldsymbol{y}|\boldsymbol{x}}} \right)$, we use the assumption $p\left( {\boldsymbol{x}} \right) = \frac{1}{B}$ again, and we have
\begin{equation}
\label{bound-3}
\begin{aligned}
H\left( {{\boldsymbol{y}|\boldsymbol{x}}} \right) & \approx \hat H\left( {{\boldsymbol{y}|\boldsymbol{x}}} \right) \\
& = \frac{1}{B}\sum\limits_{m = 1}^M {\sum\limits_{i = 1}^B {H\left( {p\left( {y_m^{\left( i \right)}|{\boldsymbol{x}^{\left( i \right)}}} \right)} \right)} }.
\end{aligned}
\end{equation}
Therefore, by exploiting $I\left( {{\boldsymbol{x}};{\boldsymbol{y}}} \right) = H\left( \boldsymbol{y} \right) - H\left( {{\boldsymbol{y}|\boldsymbol{x}}} \right)$, we have
\begin{equation}
\label{Changeable IB-4}
I\left( {{\boldsymbol{x}};{\boldsymbol{y}}} \right) \approx {{\hat I}_{{\boldsymbol{x}},{\boldsymbol{y}}}} = \hat H\left( {\boldsymbol{y}} \right) - \hat H\left( {{\boldsymbol{y}|\boldsymbol{x}}} \right).
\end{equation}
Similar to (\ref{bound-2}), (\ref{bound-3}). and (\ref{Changeable IB-4}), we can estimate $I\left( {\boldsymbol{x};\boldsymbol{\hat y}} \right)$ as
\begin{equation}
\label{Changeable IB-5}
I\left( {\boldsymbol{x};\boldsymbol{\hat y}} \right) \approx {{\hat I}_{\boldsymbol{x},{\boldsymbol{\hat y}}}} = \hat H\left( {{\boldsymbol{\hat y}}} \right) - \hat H\left( {{\boldsymbol{\hat y}}|x} \right).
\end{equation}
Given (\ref{Changeable IB-4}) and (\ref{Changeable IB-5}) we can obtain the upper bound as (\ref{theorem_betamax_1}). This completes the proof.
\end{proof}
From Theorem \ref{th:1}, we can observe that both $p\left( {{\boldsymbol{y}|\boldsymbol{x}}} \right)$ and $p\left( {{\boldsymbol{\hat y}}|\boldsymbol{x}} \right)$ affect  ${\beta _{\max }}$, which can be calculated via (\ref{methods-2}) and (\ref{JSAC-4}) in our designed system. 

To ensure the relevance between the source $\boldsymbol{x}$ and the codeword $\boldsymbol{y}$, we further add the approximated upper bound on $\beta$, ${\beta _{\max }}\left[ w \right]$ on the basis of PID algorithm to constrain the range of $\beta$. Note that the optimal value of $\beta$ to minimize MSE may not be $0$ since we jointly optimize the transmission rate and the distortion of JSCC. In the ideal case, the transmission distortion 
can be reduced close to 0 under a certain compression ratio. Therefore, we treat the MSE between the original image in the validation set and the corresponding reconstructed image at the $w$-th epoch as $e\left[ w \right]$. Then, by applying the PID algorithm, $\beta$ will change in the direction of reducing MSE. The proposed formula of adaptive $\beta$ at the $w$-th epoch is
\begin{equation}
\label{Changeable IB-2}
\begin{aligned}
{\beta _{{\rm{adp}}}}\left[ w \right] = & {\beta _{\max }}\left[ w \right] + {K_p}{\rm{MSE}}\left[ w \right] - {K_i}\sum\limits_{k = 0}^{w - 1} {{\rm{MSE}}\left[ k \right]}\\
& - {K_d}\left( {{\rm{MSE}}\left[ w \right] - {\rm{MSE}}\left[ {w - 1} \right]} \right),
\end{aligned}
\end{equation}
where ${\beta _{\max }}\left[ w \right] = \frac{{{{\hat I}_{\boldsymbol{x},{\boldsymbol{\hat y}}}}\left[ w \right]}}{{{{\hat I}_{{\boldsymbol{x}},{\boldsymbol{y}}}}\left[ w \right]}}$, ${{{\hat I}_{{{\boldsymbol{x}},{\boldsymbol{y}}}}}\left[ w \right]}$ and ${{{\hat I}_{\boldsymbol{x},{\boldsymbol{\hat y}}}}\left[ w \right]}$ are ${{{\hat I}_{{\boldsymbol{x}},{\boldsymbol{y}}}}}$ and ${{{\hat I}_{\boldsymbol{x},{\boldsymbol{\hat y}}}}}$ at the $w$-th epoch, respectively, ${\rm{MSE}}\left[ w \right]$ is the average MSE at the $w$-th epoch and it is expressed as
\begin{equation}
\label{MSE}
{\rm{MSE}}\left[ w \right] = \frac{1}{V}\frac{1}{N}\sum\limits_{i = 1}^V {\sum\limits_{j = 1}^N {{{\left( {x_j^{\left( i \right)}\left[ w \right] - \hat x_j^{\left( i \right)}\left[ w \right]} \right)}^2}} } ,
\end{equation}
with $V$ being the number of the images in the validation set, ${x_j^{\left( i \right)}\left[ k \right]}$ being the $j$-th pixel in the $i$-th transmitted image ${\boldsymbol{x}^{\left( i \right)}}$ recovered at the $k$-th epoch, and ${\hat x_j^{\left( i \right)}\left[ k \right]}$ being the corresponding pixel in the corresponding reconstructed image ${{{\boldsymbol{\hat x}}}^{\left( i \right)}}$ at the $w$-th epoch.

After training, $I\left( {\boldsymbol{x};\boldsymbol{\hat y}} \right)$ and $I\left( {{\boldsymbol{x}};{\boldsymbol{y}}} \right)$ 
slightly fluctuates in a small range, and the balance between them is nearly fixed. $\beta$ should converge to a certain minimal value. In general, the minimal value of $\beta$ is larger than $0$ since $\beta = 0$ means ignoring the compression term $I\left( {{\boldsymbol{x}};{\boldsymbol{y}}} \right)$. Therefore, we constrain $\beta$ larger than a minimum value, ${\beta _{\min }}\left( { > 0} \right)$. Then, $\beta$ at $w$-th epoch can be expressed as:
\begin{equation}
\label{JSAC-5}
\beta \left[ w \right] = {\rm{clamp}}\left( {{\beta _{{\rm{adp}}}}\left[ w \right],{\beta _{\min }},{\beta _{\max }}\left[ w \right]} \right),
\end{equation}
where ${\rm{clamp}}\left( {x,\min ,\max } \right)$ represents clamping $\boldsymbol{x}$ between $\rm{min} $ and $\rm{max}$ ($\min  \le \max $). 

\begin{algorithm}[t]
\caption{Adaptive IB Algorithm}
\begin{algorithmic}[1]
\renewcommand{\algorithmicrequire}{\textbf{Input:}}
\REQUIRE
Encoder ${p_{\boldsymbol{\varphi}} }\left( {{\boldsymbol{y}|\boldsymbol{x}}} \right)$; Decoder ${p_{\boldsymbol{\theta}} }\left( {{\boldsymbol{\hat x}}|{\boldsymbol{\hat y}}} \right)$; MSE at $(w-1)$-th epoch and $w$-th epoch: ${\rm{MSE}}[w]$ and ${\rm{MSE}}[w-1]$, Coefficients ${K_p}$, ${K_i}$ and ${K_d}$; Minimal value of hyperparameter $\beta : {\beta _{\min }}$.
\renewcommand{\algorithmicrequire}{\textbf{Output:}}
\REQUIRE
The hyperparameter used in $w$-th epoch: $\beta \left[ w \right]$.
\renewcommand{\algorithmicrequire}{\textbf{Initialization:}}
\REQUIRE
$I\left[ 0 \right] = 0$; ${\rm{MSE}}\left[ 0 \right] = 0$.
\STATE $P\left[ w \right] \leftarrow {K_p}{\rm{MSE}}\left[ {\rm{w}} \right]$;
\STATE $I\left[ w \right] \leftarrow I\left[ {w - 1} \right] + {K_i}{\rm{MSE}}\left[ {\rm{w}} \right]$;
\STATE $D\left[ w \right] \leftarrow {K_d}\left( {{\rm{MSE}}\left[ {\rm{w}} \right]{\rm{ - MSE}}\left[ {{\rm{w - 1}}} \right]} \right)$;
\STATE Calculate ${\beta _{\max }}\left[ w \right]$ according to (\ref{theorem_betamax_1});
\STATE ${\beta _{{\rm{adp}}}}\left[ w \right] \leftarrow {\beta _{\max }}[w] + P\left[ w \right] - I\left[ w \right] - D\left[ w \right]$;
\STATE $\beta \left[ w \right] \leftarrow {\rm{clamp}}\left( {{\beta _{{\rm{adp}}}}\left[ w \right],{\beta _{\min }},{\beta _{\max }}\left[ w \right]} \right)$;
\end{algorithmic}\label{algorithm1}
\end{algorithm}

\begin{algorithm}[t]
\caption{AIB-JSCC}
\begin{algorithmic}[1]
\renewcommand{\algorithmicrequire}{\textbf{Input:}}
\REQUIRE
Dataset($\mathcal{X}$) to be compressed; Channel error probability $\boldsymbol{\varepsilon}$; Hyperparameter $\beta $.
\renewcommand{\algorithmicrequire}{\textbf{Output:}}
\REQUIRE
Learned encoder ${p_{\boldsymbol{\varphi}} }\left( {{\boldsymbol{y}|\boldsymbol{x}}} \right)$ and decoder ${{p_{\boldsymbol{\theta}} }\left( {{\boldsymbol{\hat x}}|{\boldsymbol{\hat y}}} \right)}$.
\STATE Initialize the parameters of encoder ${p_{\boldsymbol{\varphi}}}\left( {{\boldsymbol{y}|\boldsymbol{x}}} \right)$, the parameters of decoder ${{p_{\boldsymbol{\theta}}}\left( {{\boldsymbol{\hat x}}|{\boldsymbol{\hat y}}} \right)}$; $i = 1$.
\WHILE{not converge}
  \STATE Sample $B$ samples from Dataset: $\boldsymbol{x} \sim p\left( {\boldsymbol{x}} \right)$;
  \STATE Sample a codeword $\boldsymbol{y} \sim {p_{\boldsymbol{\varphi}} }\left( {{\boldsymbol{y}|\boldsymbol{x}}} \right)$ for each $\boldsymbol{x}$;
  \STATE Sample $K$ noisy codewords ${\boldsymbol{\hat y}} \sim p\left( {{\boldsymbol{\hat y}}|{\boldsymbol{x}};{\boldsymbol{\varphi}} ,{\boldsymbol{\varepsilon}} } \right)$ for each $\boldsymbol{x}$;
 \STATE Calculate ${{\hat I}_{{\rm{VL}}}}\left( {\boldsymbol{x},{\boldsymbol{\hat y}};{\boldsymbol{\varphi}} ,{\boldsymbol{\theta}} ,{\boldsymbol{\varepsilon}} } \right)$ according to (\ref{VIMCO});
  \STATE Calculate ${{\hat I}_{{\rm{CLUB}}}}\left( {{\boldsymbol{x}},{\boldsymbol{y}};{\boldsymbol{\varphi}} } \right)$ according to (\ref{model-B-6});
  \STATE Update $\boldsymbol{\varphi}$ and ${\boldsymbol{\theta}}  $ according to (\ref{loss});
  \IF{an epoch of training finishes}
      \STATE Calculate ${\rm{MSE}}\left[ {\rm{i}} \right]$ according to (\ref{MSE});
      \STATE Update $\beta \left[ i \right]$ according to Algorithm \ref{algorithm1};
   \ENDIF
   \STATE $i \leftarrow i + 1$;
\ENDWHILE
\end{algorithmic}\label{algorithm2}
\end{algorithm}

From (\ref{methods-5}), the importance of $I\left( {{\boldsymbol{x}};{\boldsymbol{y}}} \right)$ in the loss function decreases as $\beta$ increases. At the beginning, $\boldsymbol{y}$ contains abundant redundancy due to imperfect map from the source information to the transmitted codewords. Therefore, in the initial stages, we must set a relatively large $\beta$ to squeeze more redundant information in $\boldsymbol{y}$. As the training processes, the value of $\beta$ should decrease since the corresponding redundancy information in $\boldsymbol{y}$ gradually decreases. The value of $\beta$ will finally converge to a constant when the proposed AIB-JSCC achieves the optimal trade-off between the reconstruction quality and the compression ratio. Therefore, we adjust the coefficients ${K_p}$, ${K_i}$ and ${K_d}$ to let $\beta$ gradually decrease from its upper bound as the training processes. The adaptive IB algorithm is summarized in Algorithm \ref{algorithm1}. 

\subsection{Training Process of AIB-JSCC}
The architecture of the AIB-JSCC system is shown in Fig. \ref{fig:AIB-JSCC}. The encoder first extracts the information of the input image as a feature map according to a feature extractor which consists of convolutional neural networks (CNN) or fully connected (FC) layer. Then, to control the length of $\boldsymbol{y}$, an FC layer is used to turn the feature map into a $M$-dimensional vector ${f_{\boldsymbol{\varphi}} }\left( {\boldsymbol{x}} \right)$. The codeword $\boldsymbol{y}$ is sampled according to $y \sim {\rm{Bern}}\left( {{f_{\boldsymbol{\varphi}} }\left( {\boldsymbol{x}} \right)} \right)$. At the receiver, the noisy codeword $\boldsymbol{\hat y}$ is first passed into an FC layer and then reshaped into a feature map. The feature map is upsampled to the same dimension as $\boldsymbol{x}$ to obtain ${{g_{\boldsymbol{\theta}} }\left( {{\boldsymbol{\hat y}}} \right)}$. The recovered image $\boldsymbol{\hat x}$ is generated according to ${\boldsymbol{\hat x}} \sim N\left( {{g_{\boldsymbol{\theta}} }\left( {{\boldsymbol{\hat y}}} \right),\boldsymbol{I}} \right)$. At the output layer of the decoder, an activation function is used to transform the pixel values in ${{\boldsymbol{\hat x}}}$ to $\left[ {0,1} \right]$. We then multiply ${{\boldsymbol{\hat x}}}$ by 255 and round the resulting values to ensure that the pixel values are discrete and fall between $\left[ {0,255} \right]$. (\ref{loss}) is used as the loss function to train the encoder and the decoder for image transmission jointly. We use the mini-batch gradient descent method\cite{bottou2010large} to optimize the parameters. To guarantee that each image in the batch has an equal probability of being selected for updating the parameters, we have $p\left( {\boldsymbol{x}} \right) = \frac{1}{B}$. Then, the distribution of $\boldsymbol{y}$ can be obtained shown in (\ref{methods-2}), and the distribution of $\boldsymbol{\hat y}$ can be obtained shown in (\ref{JSAC-4}). We sample one codeword $\boldsymbol{y}$ for each $\boldsymbol{x}$, and have $B$ pairs $\left\{ {\left( {{\boldsymbol{x}^{\left( i \right)}},{\boldsymbol{y}^{\left( i \right)}}} \right)} \right\}_{i = 1}^B$. Then, the transmission rate, ${{\hat I}_{{\rm{CLUB}}}}\left( {{\boldsymbol{x}},{\boldsymbol{y}};{\boldsymbol{\varphi}} } \right)$ is calculated based on (\ref{model-B-6}). To calculate ${{\hat I}_{{\rm{VL}}}}\left( {\boldsymbol{x},{\boldsymbol{\hat y}};{\boldsymbol{\varphi}} ,{\boldsymbol{\theta}} ,{\boldsymbol{\varepsilon}} } \right)$, we further sample $K$ noisy codewords $\boldsymbol{\hat y}$ for each $\boldsymbol{x}$, and the total number of $\boldsymbol{\hat y}$ is $B \times K$. According to (\ref{VIMCO}), the distortion term ${{\hat I}_{{\rm{VL}}}}\left( {\boldsymbol{x},{\boldsymbol{\hat y}};{\boldsymbol{\varphi}} ,{\boldsymbol{\theta}} ,{\boldsymbol{\varepsilon}} } \right)$ can be calculated. Finally, $\boldsymbol{\varphi}$ and $\boldsymbol{\theta}$ are updated according to (\ref{loss}). At the end of each epoch, Algorithm \ref{algorithm1} is applied to update $\beta$. The coefficients  ${K_p}$, ${K_i}$, and ${K_d}$ are adjusted to obtain proper $\beta$. 
The updated $\beta $ is then used in the loss function of the next epoch. The model with the lowest average MSE on valid dataset during training is stored. After training, the AIB-JSCC system can reduce the transmitted data by minimizing ${{\hat I}_{{\rm{CLUB}}}}\left( {{\boldsymbol{x}},{\boldsymbol{y}};{\boldsymbol{\varphi}} } \right)$ while improving the reconstruction quality by maximizing ${{\hat I}_{{\rm{VL}}}}\left( {\boldsymbol{x},{\boldsymbol{\hat y}};{\boldsymbol{\varphi}},{\boldsymbol{\theta}},{\boldsymbol{\varepsilon}} } \right)$. The whole training procedure of the proposed AIB-JSCC is summarized in Algorithm \ref{algorithm2}.
\begin{table}[t!]
\setlength\tabcolsep{3pt}
\renewcommand\arraystretch{1.4}
\centering
\caption{\label{tab:Parameters}System parameters.}
\begin{tabular}{|p{2cm}|c|c|c|c|}
\hline
{\textbf{Parameters}} & \multicolumn{4}{c|}{\textbf{Value}}\\
\hline
Datasets & MNIST & CIFAR10 & SVHN & Omniglot \\
\hline
\tabincell{l}{Codewords \\ length $M$} & $100$ & \tabincell{c}{400,450,\\ 500,550,600} & $500$ & $200$ \\
\hline
\multirow{2}*{\tabincell{l}{Channel error \\ probability $\boldsymbol{\varepsilon}$} }& \multicolumn{4}{c|}{Single-channel: $0.1, 0.2, 0.3, 0.4$}\\
\cline{2-5}
~ & \multicolumn{4}{c|}{Parallel-channel: shown as (\ref{JSAC-6})}\\
\hline
\tabincell{l}{Subchannel\\
number $P$} & \multicolumn{4}{c|}{$2, 4, 5$}\\
\hline
\tabincell{l}{$\beta$ ( ${\beta _{\min }}$ )} & \multicolumn{3}{c|}{$0.01$} & $0.001$\\
\hline
${K_p}$& \multicolumn{4}{c|}{$0.001$}\\
\hline
${K_i}$ & $-0.001$ & \multicolumn{3}{c|}{$-0.0001$}\\
\hline
${K_d}$& \multicolumn{4}{c|}{$-0.001$}\\
\hline
Batchisize $B $ & \multicolumn{4}{c|}{$300$}\\
\hline
Training Epoch & \multicolumn{4}{c|}{$500$}\\
\hline
Learning Rate & \multicolumn{4}{c|}{$0.001$}\\
\hline
Regularization coefficient & \multicolumn{4}{c|}{$0.0001$}\\
\hline
\end{tabular}
\end{table}

\begin{table}[t!]
\setlength\tabcolsep{5pt}
\renewcommand\arraystretch{1.4}
\centering
\caption{\label{tab:psnr}PSNR of NECST vs. IABF vs. IB-JSCC vs. AIB-JSCC.}
\begin{tabular}{|c|c|c|c|c|c|}
\hline
\multirow{3}*{\textbf{Datasets}} & \multirow{3}*{\textbf{Methods}} & \multicolumn{4}{c|}{\textbf{PSNR under different} } \\
~ & ~ & \multicolumn{4}{c|}{\textbf{ error probabilities}} \\
\cline{3-6}
~ & ~ & \textbf{0.1} & \textbf{0.2} & \textbf{0.3} & \textbf{0.4}\\
\hline
\multirow{4}*{MNIST} & NECST & 17.348 & 15.411  & 13.581 & 12.104  \\   
\cline{2-6}
~ & IABF & 17.721
 & 15.513  & 13.735 & 12.264  \\
\cline{2-6}
~ & IB-JSCC & 17.801
 & 15.724  & 13.741 & 12.408  \\
\cline{2-6}
~ & AIB-JSCC & \textbf{17.837} & \textbf{15.725} & \textbf{13.751} & \textbf{12.411} \\
\hline
\multirow{4}*{Omniglot}  & NECST & 15.017 
 & 13.955  & 12.959 & 12.1409  \\ 
\cline{2-6}
~ & IABF & 15.117
 & 13.928  & 13.039 & 12.166  \\
\cline{2-6}
~ & IB-JSCC & 15.158
 & 14.015  & 13.04 & 12.203  \\
\cline{2-6}
~ & AIB-JSCC & \textbf{15.161} & \textbf{14.03} & \textbf{13.052} & \textbf{12.213} \\
\hline
\multirow{4}*{CIFAR10} & NECST & 16.864 & 16.158 & 15.35 & 14.163  \\ 
\cline{2-6}
~ & IABF & 17.442
 & 16.391  & 15.673  & 14.219  \\
\cline{2-6}
~ & IB-JSCC & 17.455  & 16.68 & 15.792 & 14.247  \\
\cline{2-6}
~ & AIB-JSCC & \textbf{17.513} & \textbf{16.748} & \textbf{15.809} & \textbf{14.282} \\
\hline
\end{tabular}
\end{table}

\begin{figure}[t]
\centering
\includegraphics[width=0.9\columnwidth]{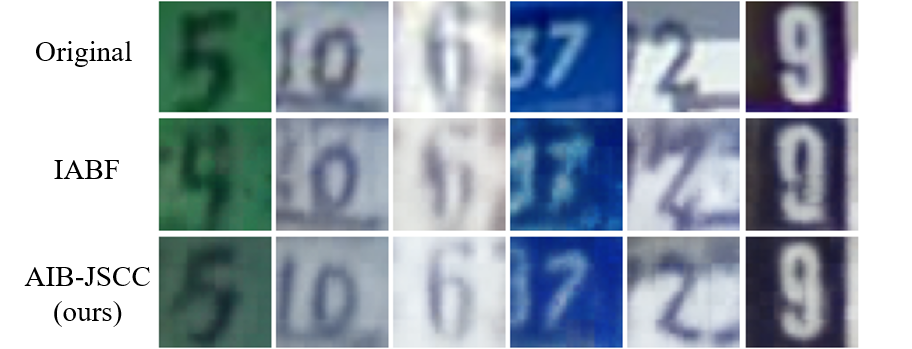}
\centering
\caption{\label{fig:reconstruction} Visual comparison between IABF and IB-JSCC.}
\end{figure}
\section{Experimental Results}
In this section, we provide extensive experiments to validate our designed system. The experiments are carried on the following datasets: MNIST\cite{lecun1998mnist}, Omniglot\cite{lake2015human}, CIFAR10\cite{krizhevsky2009learning} and street view housing numbers tsiscon (SVHN)\cite{netzer2011reading} to account for different image sizes and colors. For comparison purposes, we choose IABF and NECST in \cite{song2020infomax} and \cite{choi2019neural} and classical SSCC schemes as baselines. Specifically, for SSCC schemes, we employ three industry-standard source encoders: JPEG\cite{wallace1992jpeg}, JPEG2000\cite{rabbani2002book} and WebP\cite{2015webp},  and BPG\cite{BPG}, combined with an ideal capacity-achieving channel code (marked as “JPEG + Capacity”, “JPEG2000 + Capacity” and “WebP + Capacity”, and “BPG + Capacity”, respectively). We do not compare with LDPC coding as we are often unable to obtain valid image files after LDPC decoding. To make a fair comparison, the settings and structure of the encoder and decoder neural networks used in AIB-JSCC are the same as those used in IABF. The system parameters are shown in Table \ref{tab:Parameters}. In line with the baseline and references\cite{choi2019neural,song2020infomax}, we choose similar parameters of the neural network and the optimizer. For Monte Carlo estimation of ${{I_{{\rm{VL}}}}\left( {\boldsymbol{x};\boldsymbol{\hat y}} \right)}$, we utilize $5$ samples\cite{choi2019neural}\cite{song2020infomax}. The Monte Carlo estimates of ${{\hat I}_{{\rm{CLUB}}}}\left( {{\boldsymbol{x}}, {\boldsymbol{y}};{\boldsymbol{\varphi}} } \right)$, ${\rm{H}}\left( {\boldsymbol{y}} \right)$, ${\rm{H}}\left( {{\boldsymbol{y}}|{\boldsymbol{x}}} \right)$ and ${\rm{H}}\left( {\boldsymbol{\hat y}} \right)$ use $300$ samples per batch. We choose the best ${K_p}$ and ${K_d}$ from the set $\left\{ {{{10}^{ - 2}},{{10}^{ - 3}},{{10}^{ - 4}},{{10}^{ - 5}}} \right\}$ that provide the best performance. We use widely-used image quality metrics, MSE and peak signal-to-noise ratio (PSNR), to measure the performance of AIB-JSCC and IABF  \cite{song2020infomax}. In the single-channel scenario, we compare the reconstruction and compression ability of AIB-JSCC with baselines, and the robustness and complexity of AIB-JSCC are also discussed. In the parallel-channel scenario, we present the results of the reconstruction error, the distribution of neuron weights and the visual reconstructions to illustrate that AIB-JSCC can adaptively allocate elements to parallel channels according to their channel state information. The above experiments are implemented for the BSC. We also compare the reconstruction error of AIB-JSCC and IABF when the channel is the high-order DMC to demonstrate the effectiveness of AIB-JSCC. 

\subsection{Single-channel Scenario}

\subsubsection{Reconstruction capabilitiy}
Table \ref{tab:psnr} shows the PSNR of different schemes on various datasets under different error probabilities, where IB-JSCC stands for the degenerate AIB-JSCC with fixed $\beta$. The definition of PSNR is
\begin{equation}
\label{PSNR}
{\rm{PSNR}} = 10{\log _{10}}\left( {\frac{{{{\left( {{{\rm{2}}^n} - 1} \right)}^2}}}{{{\rm{mse}}\left( {{\boldsymbol{x}},{\boldsymbol{\hat x}}} \right)}}} \right),
\end{equation}
where $n$ is the number of bits that each image pixel uses, ${\rm{mse}}\left( {{\boldsymbol{x}},{\boldsymbol{\hat x}}} \right)$ is MSE between ${\boldsymbol{x}}$ and ${{\boldsymbol{\hat x}}}$. In particular, we fix the length of $\boldsymbol{y}$ and calculate the average MSE and PSNR over the test sets. From Table \ref{tab:psnr}, we can observe that AIB-JSCC is always superior to IB-JSCC and IABF in terms of MSE and PSNR, which validates the effectiveness of the proposed IB objective and the adaptive IB algorithm. From Table \ref{tab:psnr}, we can also observe that AIB-JSCC and IB-JSCC can reduce MSE and increase PSNR more on the RGB dataset CIFAR10 than on the greyscale datasets MNIST and Omniglot. This is because AIB-JSCC can extract information more precisely with the guidance of the proposed IB objective, thus recovering complex images better.

\begin{table}[t]
\renewcommand\arraystretch{1.4}
\caption{\label{tab:classification}Classification accuracy of images recovered by IABF and AIB-JSCC}
\begin{center}
\begin{tabular}{|c|c|c|c|c|c|}
\hline
\multirow{3}*{\textbf{Classifiers}} & \multirow{3}*{\textbf{Methods}} & \multicolumn{4}{c|}{\textbf{Acc under different} } \\
~ & ~ & \multicolumn{4}{c|}{\textbf{ error probabilities}} \\
\cline{3-6}
~ & ~ & \textbf{0.1} & \textbf{0.2} & \textbf{0.3} & \textbf{0.4}\\
\hline
\multirow{2}*{MLP} & IABF & 0.932
 & 0.817 & 0.637 & 0.331 \\
\cline{2-6}
~ & AIB-JSCC & \textbf{0.937} & \textbf{0.881} & \textbf{0.692} & \textbf{0.386} \\
\hline
\multirow{2}*{SVM} & IABF & 0.932 &	0.821 & 0.619 & 0.312 \\
\cline{2-6}
~ & AIB-JSCC & \textbf{0.942} & \textbf{0.884} & \textbf{0.694} & \textbf{0.358} \\
\hline
\multirow{2}*{DT} & IABF & 0.51 &	0.391 & 0.297 & 0.177 \\
\cline{2-6}
~ & AIB-JSCC & \textbf{0.564} & \textbf{0.469} & \textbf{0.347} & \textbf{0.2} \\
\hline
\multirow{2}*{RF} & IABF & 0.673 & 0.522 & 0.288 & 0.176 \\
\cline{2-6}
~ & AIB-JSCC & \textbf{0.708} & \textbf{0.549} & \textbf{0.308} & \textbf{0.181} \\
\hline
\end{tabular}
\end{center}
\end{table}

Figure \ref{fig:reconstruction} shows the visual reconstructions of AIB-JSCC and IABF on the SVHN dataset where ${\boldsymbol{\varepsilon}} {\rm{ = 0}}{\rm{.1}}$ and $M=500$. From Fig. \ref{fig:reconstruction}, we can observe that images recovered by AIB-JSCC are closer to original ones than those recovered by IABF, and the numbers in images recovered by AIB-JSCC can be distinguished more easily than IABF. For example, the first image recovered by AIB-JSCC can be clearly recognized as $5$ while the one recovered by IABF may be incorrectly recognized as $9$ or $4$. This implies that AIB-JSCC can preserve more semantic information than IABF. This is due to the fact that compared with IABF, AIB-JSCC can preserve useful information as well as discard useless information which may lead to semantic mistakes. In consequence, AIB-JSCC has better visual reconstruction quality.
\begin{figure}[t]
\centering
\includegraphics[width=1\columnwidth]{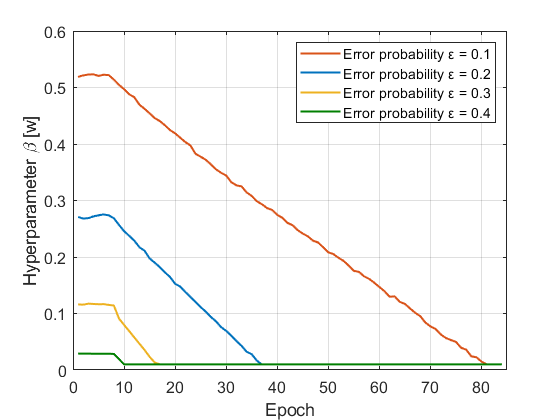}
\centering
\caption{\label{fig:beta} The value of hyperparameter $\beta \left[ w \right]$ with respect to training epoch.}
\end{figure}

\begin{figure}[t]
\centering
\includegraphics[width=1\columnwidth]{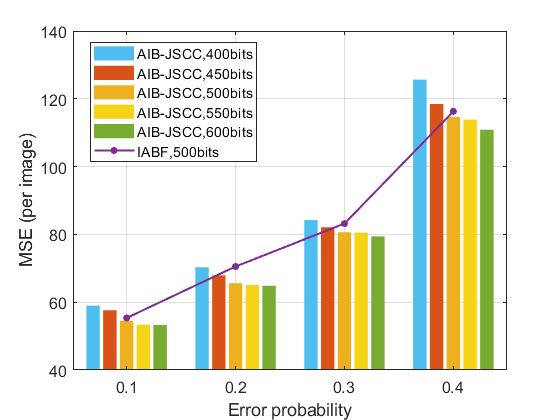}
\centering
\caption{\label{fig:compression-IABF} MSE of IABF and AIB-JSCC with different $M$.}
\end{figure}

\begin{figure}[t]
\centering
\includegraphics[width=1\columnwidth]{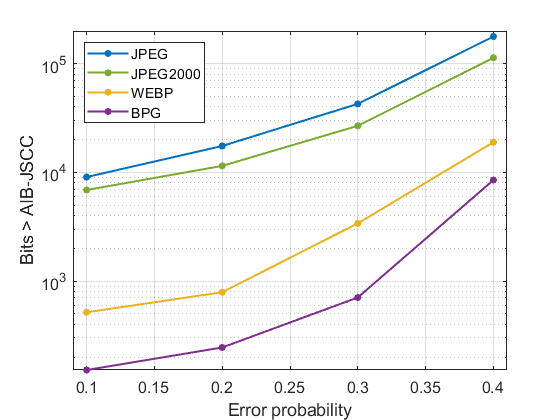}
\centering
\caption{\label{BPG-ib-jscc-sscc} The additional number of bits need by SSCC.}
\end{figure}

\begin{figure}[t]
\centering
\includegraphics[width=1\columnwidth]{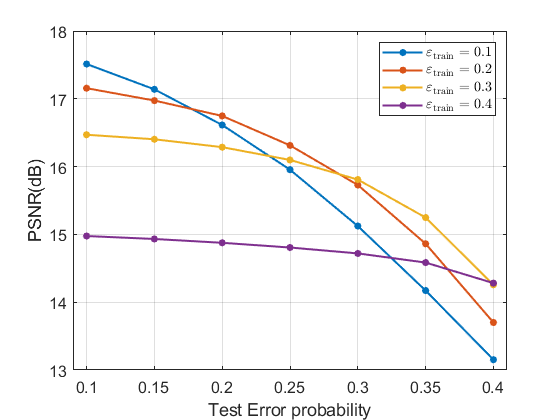}
\centering
\caption{\label{train_test_various} The PSNR of AIB-JSCC with various train and test error probabilities.}
\end{figure}

Table \ref{tab:classification} shows the classification accuracy (Acc) of the images reconstructed by IABF and AIB-JSCC with respect to different error probabilities. In particular, 4 different classifiers, multilayer perceptron (MLP), support vector machines (SVM), decision trees (DT) and random forests (RF) are trained with the raw MNIST train set, and tested with the images reconstructed by IABF and AIB-JSCC from the MNIST test set where ${\boldsymbol{\varepsilon}} {\rm{ = 0}}{\rm{.1}}$. From Table \ref{tab:classification}, we can observe that the classification accuracy of the images recovered by AIB-JSCC is always higher than IABF over different error probabilities. This implies that AIB-JSCC can preserve more semantic information useful for downstream task. This is because the proposed IB objective preserves information as well as discards information. Hence, AIB-JSCC can extract information more precisely.

Figure \ref{fig:beta} shows the trend of AIB-JSCC's hyperparameter $\beta \left[ w \right]$ with respect to training epoch on CIFAR10 dataset under different error probabilities. From Fig. \ref{fig:beta}, we can observe that $\beta \left[ w \right]$ gradually decreases to a minimal value ${\beta _{\min }}$ when training processes. This is due to the fact that when the training processes, $I\left( {\boldsymbol{x};{\boldsymbol{\hat y}}} \right)$ increases and $I\left( {{{\boldsymbol{x}};{\boldsymbol{y}}}} \right)$ decreases, and to keep balance between $I\left( {\boldsymbol{x};{\boldsymbol{\hat y}}} \right)$ and $I\left( {{{\boldsymbol{x}};{\boldsymbol{y}}}} \right)$, the proposed adaptive IB algorithm decreases $\beta \left[ w \right]$ to reduce the proportion of $I\left( {\boldsymbol{x};{\boldsymbol{\hat y}}} \right)$ in the loss function. From Fig. \ref{fig:beta}, we can also observe that the value of  $\beta \left[ w \right]$ reduces as the error probability increases. This is because when the channel error probability increases, we need to add more redundancy to the transmitted codeword ${\boldsymbol{y}}$. In AIB-JSCC, this is achieved by increasing the distortion term $I\left( {{\boldsymbol{x}};{\boldsymbol{\hat y}}} \right)$. Since there is a Markov chain relationship ${\boldsymbol{x}} \to {\boldsymbol{y}} \to {\boldsymbol{\hat y}} \to {\boldsymbol{\hat x}}$, $I\left( {{\boldsymbol{x}};{\boldsymbol{\hat y}}} \right) \le I\left( {{\boldsymbol{x}};{\boldsymbol{y}}} \right) \le H\left( {\boldsymbol{y}} \right)$, and so maximizing $I\left( {{\boldsymbol{x}};{\boldsymbol{\hat y}}} \right)$ can essentially increase $I\left( {{\boldsymbol{x}};{\boldsymbol{y}}} \right)$ and $H\left( {\boldsymbol{y}} \right)$, thus increasing the redundancy in ${\boldsymbol{y}}$. This implies that the proposed adaptive IB algorithm is able to adjust $\beta \left[ w \right]$ according to $I\left( {\boldsymbol{x};\boldsymbol{\hat y}} \right)$, $I\left( {\boldsymbol{x}; \boldsymbol{y}} \right)$ and the  error probability.
\begin{figure*}[t!]
\centering 
\setlength{\abovecaptionskip}{0cm}
\setlength{\belowcaptionskip}{-0.5cm}
\subfigbottomskip=7pt 
\subfigcapskip=0pt 
\subfigure[${\boldsymbol{\varepsilon}}  = 0.1$.]{
\label{valid_loss1}
\includegraphics[width=.46\linewidth]{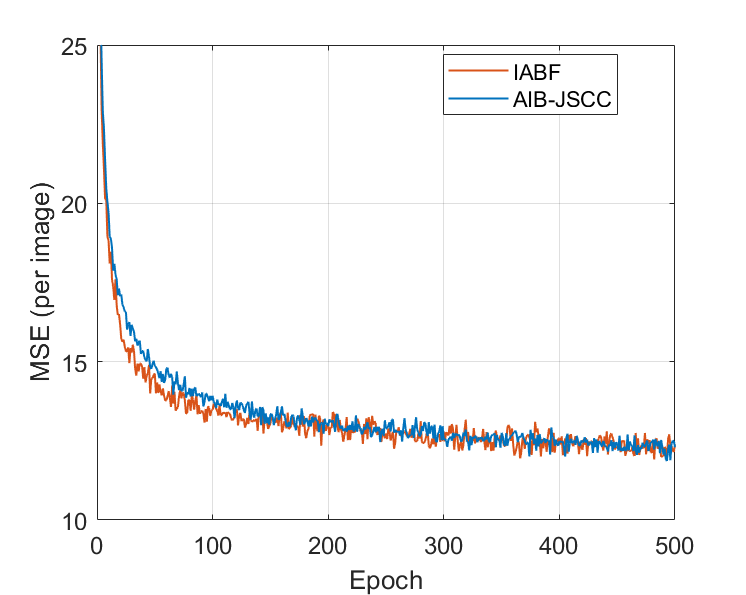}}\hspace{4pt}
\subfigure[${\boldsymbol{\varepsilon}}  = 0.2$.]{
\label{valid_loss2}
\includegraphics[width=.46\linewidth]{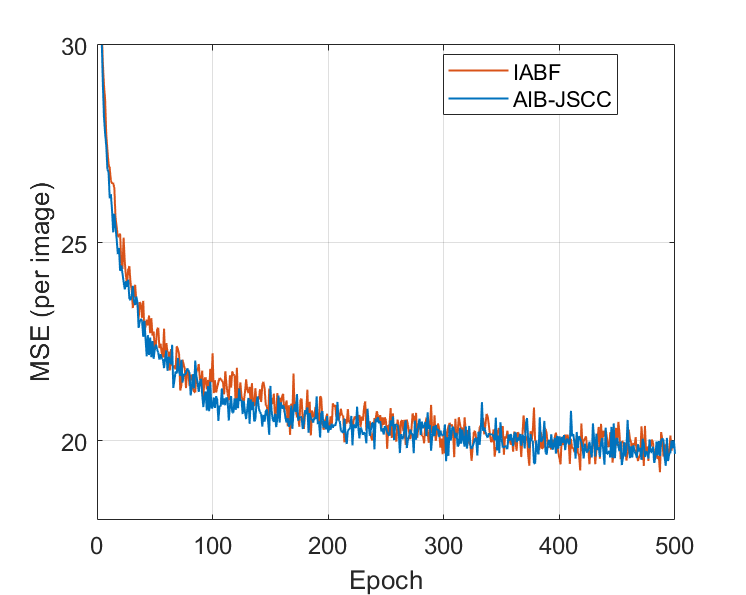}}\vspace{-1pt}
\subfigure[${\boldsymbol{\varepsilon}}  = 0.3$.]{
\label{valid_loss3}
\includegraphics[width=.46\linewidth]{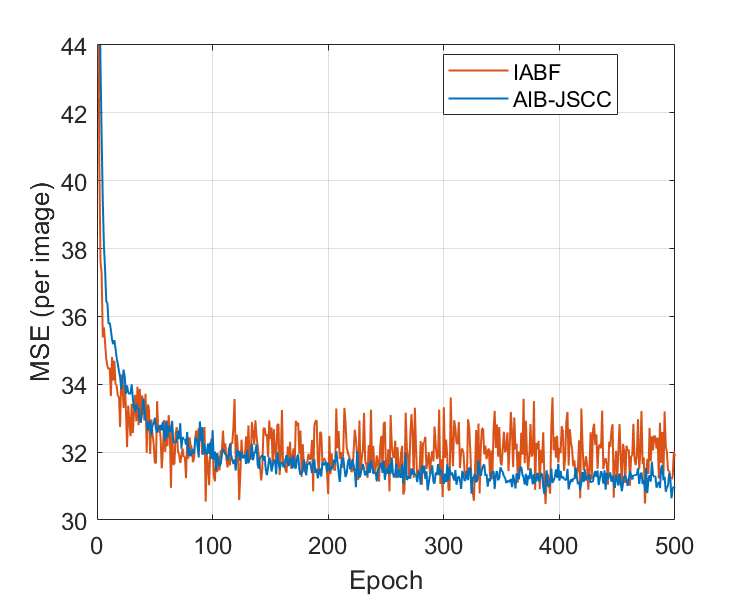}}\hspace{4pt}
\subfigure[${\boldsymbol{\varepsilon}}  = 0.4$.]{
\label{valid_loss4}
\includegraphics[width=.46\linewidth]{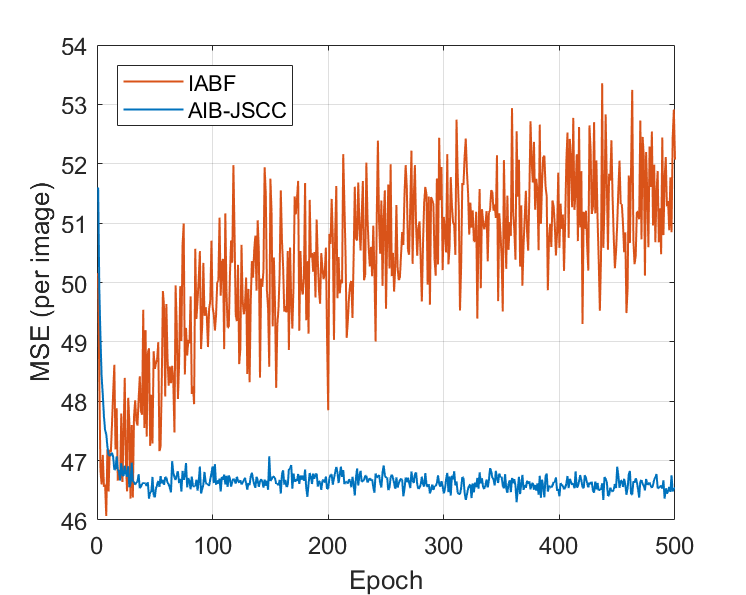}}
\caption{MSE of IABF and AIB-JSCC on MNIST dataset. The error is calculated on validation set during training.}
\label{fig:valid_loss}
\end{figure*}

\begin{table}[t!]
\renewcommand\arraystretch{1.4}
\centering
\caption{\label{tab:complexity}The number of network parameters and the inference time of IABF and AIB-JSCC.}
\begin{tabular}{|l|l|c|c|}
\hline
\textbf{Datasets} & \textbf{Methods} & \makecell[c]{\textbf{parameters } \\ \textbf{number \textbf{($ \times {\rm{1}}{{\rm{0}}^{\rm{5}}}$)}}} & \makecell[c]{\textbf{Inference } \\ \textbf{time (ms)}} \\
\hline
\multirow{2}*{MNIST} & IABF & 11.86 & 0.4  \\
\cline{2-4}
~ & AIB-JSCC & \textbf{11.345} & 0.4 \\
\hline
\multirow{2}*{Omniglot} & IABF & 13.36 & 0.743  \\
\cline{2-4}
~ & AIB-JSCC & \textbf{12.345} & \textbf{0.372} \\
\hline
\multirow{2}*{CIFAR10} & IABF & 5.679 & 1  \\
\cline{2-4}
~ & AIB-JSCC & \textbf{3.164} & \textbf{0.8} \\
\hline
\end{tabular}
\end{table}

\subsubsection{Compression Capability} 
In this section, we denote the length of $\boldsymbol{y}$, $M$, used by AIB-JSCC as the number of bits in order to compare the compression capability with other baselines. Note that $M$ bits is an upper bound on the transmission rate $I\left( {{\boldsymbol{x}};{\boldsymbol{y}}} \right)$. Figure \ref{fig:compression-IABF} shows the reconstruction MSE of IABF and AIB-JSCC with different $M$ on CIFAR10 under different error probabilities. In particular, to guarantee fairness, we use the results of IABF present in \cite{song2020infomax} in order to prevent the performance reduction caused by improper hyperparameter selection. From Fig. \ref{fig:compression-IABF}, we can observe that to obtain similar MSE, AIB-JSCC requires 15, 100, 70, 20 fewer bits than IABF when ${\boldsymbol{\varepsilon}}  = 0.1,0.2,0.3,0.4$. This implies that AIB-JSCC can reduce more than $20\%$ transmission rate compared with IABF. The $20\%$ gain stems from the fact that AIB-JSCC simultaneously minimizes the distortion and the transmission rate thus reducing the transmission rate to achieve a similar reconstruction error.

Figure \ref{BPG-ib-jscc-sscc} shows the additional number of bits that SSCC schemes need to achieve similar MSE on SVHN, compared with AIB-JSCC. From Fig. \ref{BPG-ib-jscc-sscc}, we can observe that SSCC schemes need more bits than AIB-JSCC at all datasets and error probabilities. We can also observe that although BPG + Capacity needs fewer bits than the other three SSCC schemes, BPG + Capacity still needs more bits than AIB-JSCC for all datasets and error probabilities. When the error probability $\boldsymbol{\varepsilon}$ increases, the additional required number of bits will increase. When ${\boldsymbol{\varepsilon}}  = 0.4$, AIB-JSCC only needs around $4\% $ JPEG needs. The $4\% $ gains stem from the fact that SSCC schemes are designed to be optimized for squared error with hand-selected constraints\cite{NEURIPS2018_53edebc5, 9156817, NEURIPS2020_8a50bae2} while AIB-JSCC jointly trains the encoder and decoder by maximizing $I\left( {\boldsymbol{x};{\boldsymbol{\hat y}}} \right)$ and minimizing $I\left( {{\boldsymbol{x}},{\boldsymbol{y}}} \right)$ thus preserving information precisely with lower transmission rate. 

\begin{figure*}[t!]
\centering 
\setlength{\abovecaptionskip}{0cm}
\setlength{\belowcaptionskip}{-0.5cm}
\subfigbottomskip=7pt 
\subfigcapskip=0pt 
\subfigure[${\boldsymbol{\varepsilon}} _{{\rm{test}}}  = 0.1$.]{
\label{TSNE-method1}
\includegraphics[width=.45\linewidth]{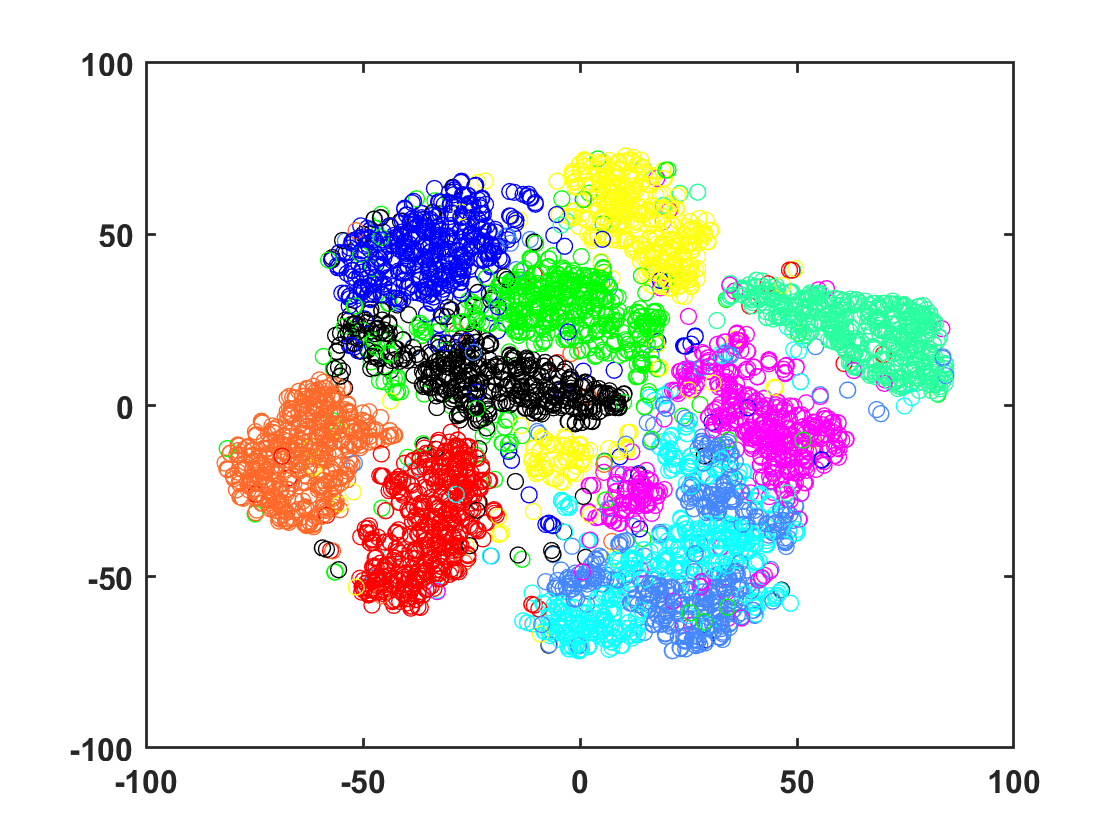}}\hspace{3pt}
\subfigure[${\boldsymbol{\varepsilon}} _{{\rm{test}}}  = 0.2$.]{
\label{TSNE-method2}
\includegraphics[width=.45\linewidth]{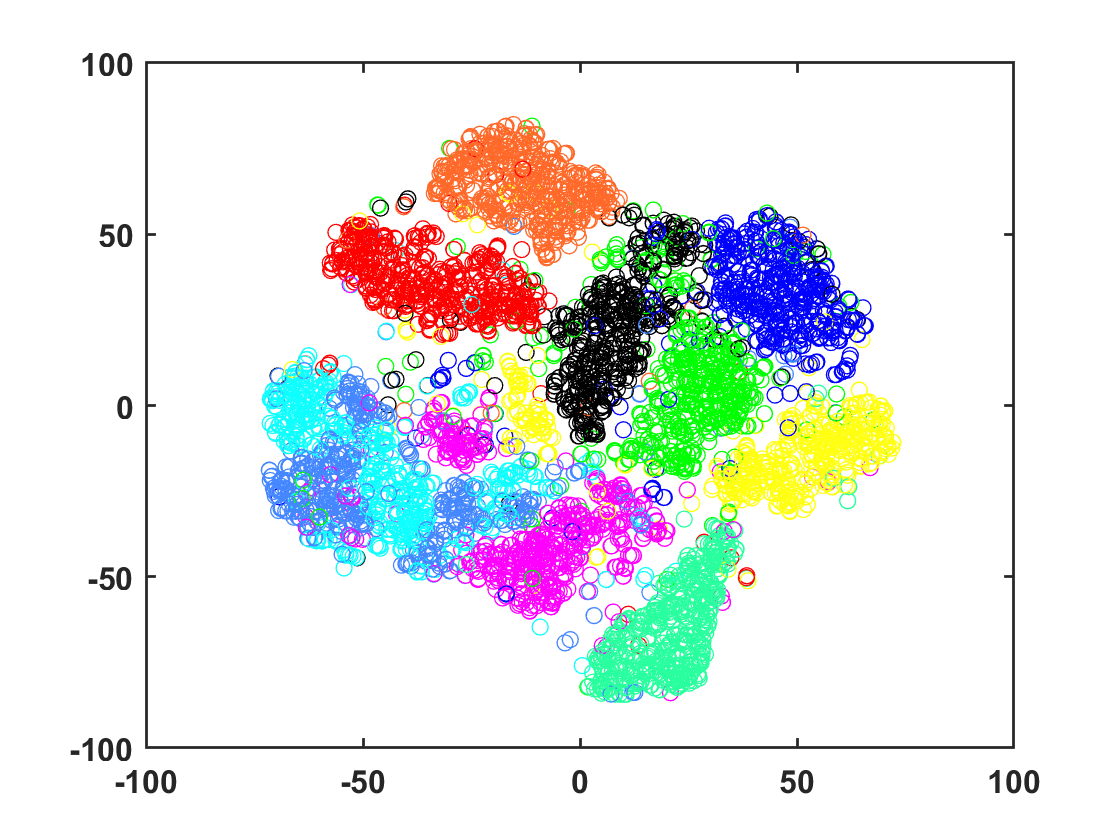}}\vspace{1pt}
\subfigure[${\boldsymbol{\varepsilon}} _{{\rm{test}}}  = 0.3$.]{
\label{TSNE-method3}
\includegraphics[width=.45\linewidth]{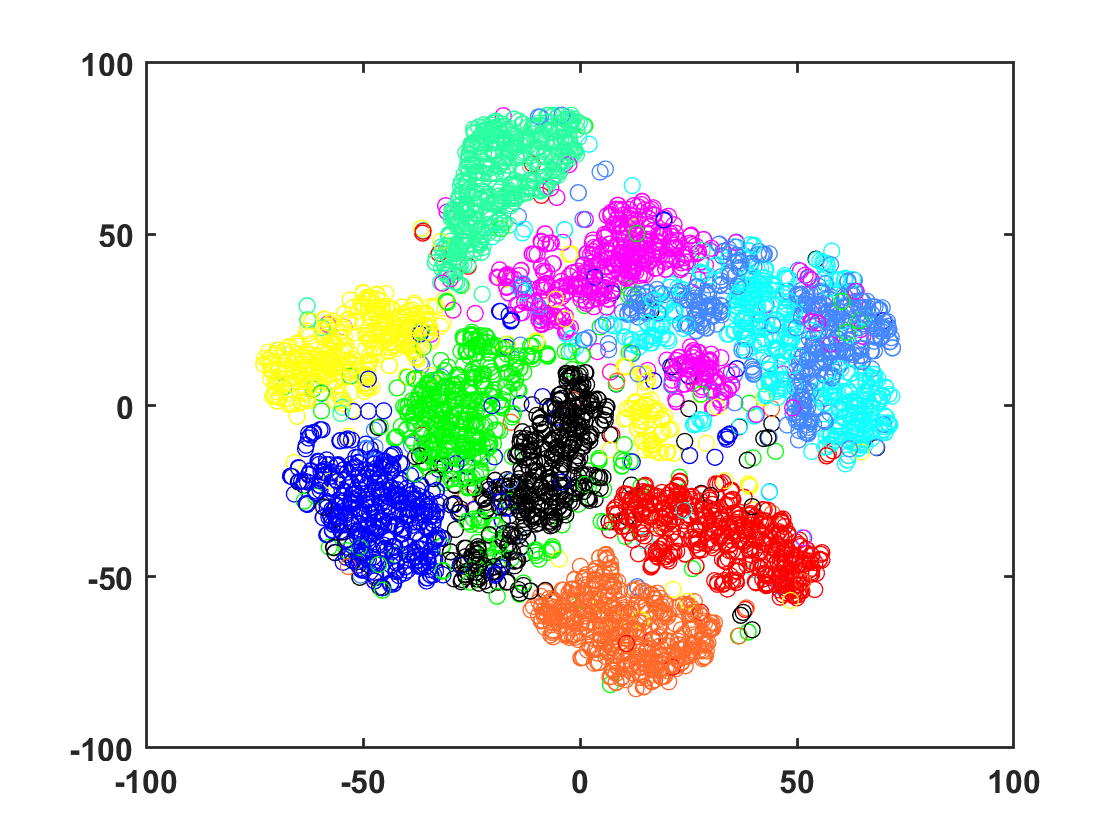}}\hspace{3pt}
\subfigure[${\boldsymbol{\varepsilon}} _{{\rm{test}}}  = 0.4$.]{
\label{TSNE-method4}
\includegraphics[width=.46\linewidth]{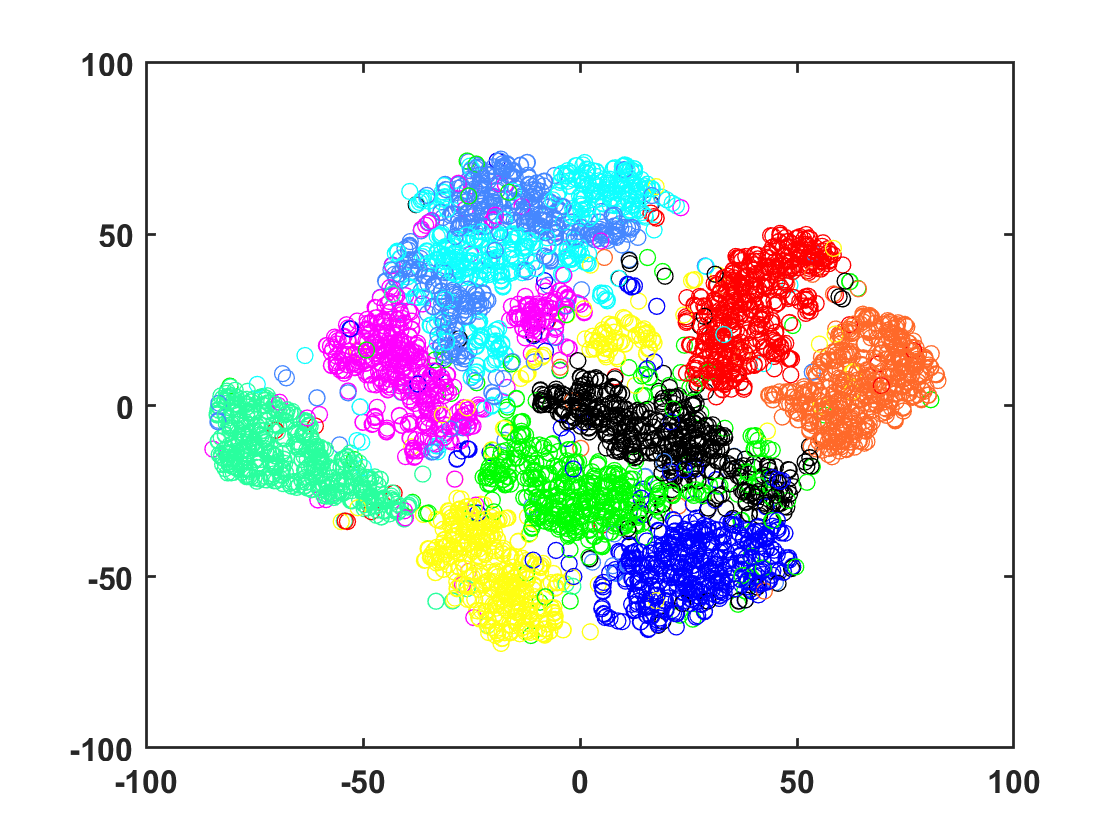}}
\caption{t-SNE visualization of codewords extracted by AIB-JSCC for test set of MNIST dataset. The network is trained with error probability ${\boldsymbol{\varepsilon}}  = 0.1$ and tested with different test error probabilities ${{\boldsymbol{\varepsilon}} _{{\rm{test}}}}$. Each color represents a different class.}
\label{fig:TSNE-method}
\end{figure*}

\begin{table}[t!]
\renewcommand\arraystretch{1.4}
\centering
\caption{\label{tab:complexity_sscc}The inference time of AIB-JSCC and SSCC.}
\begin{tabular}{|c|c|c|c|c|c|}
\hline
\textbf{Methods} & AIB-JSCC & BPG & WebP & JPEG2000 & JPEG\\
\hline
\makecell[c]{\textbf{Inference } \\ \textbf{time (ms)}} & 0.8 & 109 & 1.632 & 1.387 & 0.911\\
\hline
\end{tabular}
\end{table}

\begin{table}[t!]
\renewcommand\arraystretch{1.4}
\caption{MSE of AIB-JSCC under different parallel-channel scenarios}
\begin{center}
\begin{tabular}{|c|c|c|c|c|}
\hline
\multirow{2}*{\textbf{Scenario}} & \multirow{3}*{\textbf{Average of $\boldsymbol{\varepsilon}$}} & \multicolumn{3}{c|}{\textbf{Datasets} }\\
\cline{3-5}
~ & ~ & \textbf{MNIST} & \textbf{Omniglot} & \textbf{CIFAR10}\\
\hline
\textbf{${{\rm{2 - ch}}}$} & 0.051 & \textbf{9.616} & \textbf{23.538} & \textbf{45.069}\\
\hline
\textbf{${{\rm{4 - ch}}}$} & 0.138 & 12.398 & 27.599 & 52.834\\
\hline
\textbf{${{\rm{5 - ch}}}$} & 0.213 & 14.075 & 28.927 & 54.021 \\
\hline
\multirow{2}*{single-channel} & 0.1 & 12.902 & 23.89 & 54.464 \\
\cline{2-5}
~ & 0.2 & 20.784 & 31.092 & 64.969 \\
\hline
\end{tabular}
\label{tab:parallel-MSE}
\end{center}
\end{table}

\subsubsection{Complexity and robustness}
The most computationally costly operations in the network are the convolutions/deconvolutions and the FC layers, as they involve multiplications and additions. The computational cost of a single convolutional layer is $H \times W \times K \times K \times {C_i} \times {C_o}$, where $K$ is the filter size, ${C_o}$ is the number of output channels, ${C_i}$ is the number of input channels and $H \times W$ is the size of the feature map. The computational cost of an FC layer is $\left( {2I - 1} \right)O$, where $I$ is the input vector dimension and $O$ is the output vector dimension. Only the width and height of the feature map and the vector dimension depend on the image dimensions, and all other factors are constant and independent of the image size. Thus, the computational complexity of the proposed scheme is ${\rm O}\left( {{I_H} \times {I_W}} \right)$, where ${I_H}$ and ${I_W}$ are the width and height of the input image.

Table \ref{tab:complexity} shows the number of encoder and decoder network parameters and the inference time of IABF and AIB-JSCC. From Table \ref{tab:complexity}, we can observe that AIB-JSCC has fewer parameters and needs less inference time. Specifically, AIB-JSCC can reduce 45\% parameters on CIFAR10 and 50\% inference time on Omniglot. These gains stem from the simple network structure of AIB-JSCC, which makes the computational complexity of AIB-JSCC lower than that of IABF.

Table \ref{tab:complexity_sscc} shows the inference time of AIB-JSCC, JPEG, JPEG2000, WebP and BPG on SVHN. From Table \ref{tab:complexity_sscc}, we can observe that AIB-JSCC achieves lower inference time compared to all considered source coding schemes. Furthermore, it is worth noting that SSCC necessitates iterative channel decoding to attain optimal error correction capability \cite{lentmaier2010iterative,hagenauer1996iterative,vikalo2004iterative}. As a result, the time required by SSCC is significantly higher than that of source coding. Consequently, in comparison to practical SSCC, AIB-JSCC is expected to yield superior time savings, which are not entirely reflected in Table \ref{tab:complexity_sscc}.

Figure \ref{train_test_various} shows the PSNR of AIB-JSCC on CIFAR10 when there is an estimation error on the channel error probability. From Fig. \ref{train_test_various}, we can observe that when ${\varepsilon _{{\rm{test}}}}$ drops below ${\varepsilon _{{\rm{train}}}}$, the performance does not saturate immediately. When ${\varepsilon _{{\rm{test}}}}$ increases beyond ${\varepsilon _{{\rm{train}}}}$, AIB-JSCC exhibits a graceful degradation of the reconstruction quality. This is because AIB-JSCC uses the channel conditions in the loss function and enables the learned codewords to resist channel interference. Hence, the codewords extracted by AIB-JSCC is robust to different error probabilities.

Figure \ref{fig:valid_loss} shows the changes of validation reconstruction MSE with respect to training time steps for IABF and AIB-JSCC. From Fig. \ref{fig:valid_loss}, we can observe that the trends of IABF and AIB-JSCC are similar when error probability $\boldsymbol{\varepsilon}$ is 0.1 and 0.2. As the error probability gets larger, AIB-JSCC converges more stably than IABF. For example, in Fig. \ref{valid_loss4}, when $\boldsymbol{\varepsilon}$ is 0.4, there is severe overfitting in IABF while AIB-JSCC still converges stably. This is because AIB-JSCC avoids overfitting according to minimizing $I\left( {{{\boldsymbol{x}};{\boldsymbol{y}}}} \right)$ by neural network. Therefore, AIB-JSCC is more robust than IABF. 

Figure \ref{fig:TSNE-method} shows the 2-dimensional projections of the noisy codewords extracted from MNIST with different test error probabilities, and each color represents a number, i.e. a type in MNIST dataset. In particular, we inject noise with different error probabilities into the learned codewords extracted from MNIST and utilize t-Distributed Stochastic Neighbor Embedding (t-SNE)\cite{van2008visualizing} to project the noisy codewords into a 2-dimensional space. From Fig. \ref{fig:TSNE-method}, we can observe that noisy codewords with the same color are close to each other and well separated with other colors, and the distributions of codewords are similar under different ${{\boldsymbol{\varepsilon}} _{{\rm{test}}}}$. This is because AIB-JSCC uses the channel conditions in the loss function and enables the learned codewords to resist the channel interference. Therefore, the codewords extracted by AIB-JSCC can preserve semantic information and is robust to different error probabilities. 

\subsection{Parallel-channel Scenario}
This subsection evaluates the performance of AIB-JSCC in the parallel-channel scenario. The following three cases with different numbers of subchannels and error probabilities are considered:
\begin{equation}
\label{JSAC-6}
{\boldsymbol{\varepsilon}}  = \left\{ {\begin{array}{*{20}{c}}
{\left\{ {0.001,0.1} \right\}}&{{\rm{2 - ch}}}\\
{\left\{ {0.001,0.1,0.2,0.25} \right\}}&{{\rm{4 - ch}}}\\
{\left\{ {0.001,0.1,0.2,0.25,0.3} \right\}}&{{\rm{5 - ch}}}
\end{array}} \right.,
\end{equation}
where $P - {\rm{ch}}$ represents parallel-channel scenario with $P$ subchannels. Here, the total bandwidth is equally divided into $P$ subchannels. 

Table. \ref{tab:parallel-MSE} illustrates the reconstruction MSE over test sets. From Table \ref{tab:parallel-MSE}, we can observe that AIB-JSCC can achieve better performance in parallel-channel scenarios than that in single-channel scenarios even with smaller error probability. For instance, in the $4-\rm{ch}$ scenario, the average error probability of four subchannels is $0.138$, and the reconstruction error on MNIST and CIFAR are $12.398$ and $52.834$. In contrast, as shown in Table. \ref{tab:parallel-MSE}, in the single-channel with smaller error probability, e.g. ${\boldsymbol{\varepsilon}} {\rm{ = 0}}{\rm{.1}}$, the reconstruction error on MNIST and CIFAR are $12.902$ and $54.464$. This is because in the parallel-channel scenarios, AIB-JSCC utilizes the channel state information in the loss function and is able to transmit important elements over the subchannel with small error probability. Therefore, AIB-JSCC can dynamically allocate elements according to the error probabilities of the subchannels thus improving the reconstruction quality.

\begin{figure*}[t!]
\centering 
\setlength{\abovecaptionskip}{0cm}
\setlength{\belowcaptionskip}{-0.5cm}
\subfigbottomskip=7pt 
\subfigcapskip=0pt 
\subfigure[$2 - {\rm{ch}}$.]{
\label{a_2}
\includegraphics[width=.3\linewidth]{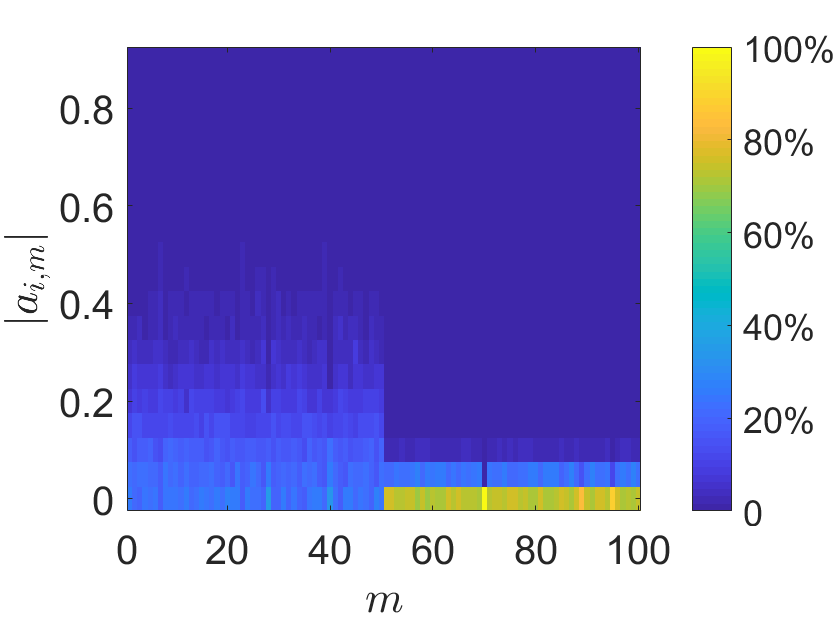}}\hspace{2pt}
\subfigure[$4 - {\rm{ch}}$.]{
\label{a_4}
\includegraphics[width=.3\linewidth]{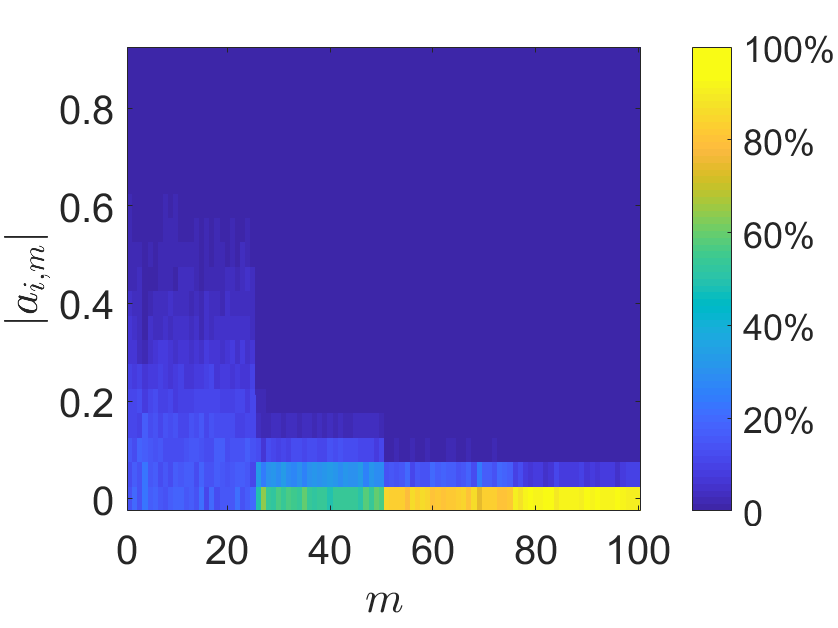}}\hspace{2pt}
\subfigure[$5 - {\rm{ch}}$.]{
\label{a_5}
\includegraphics[width=.3\linewidth]{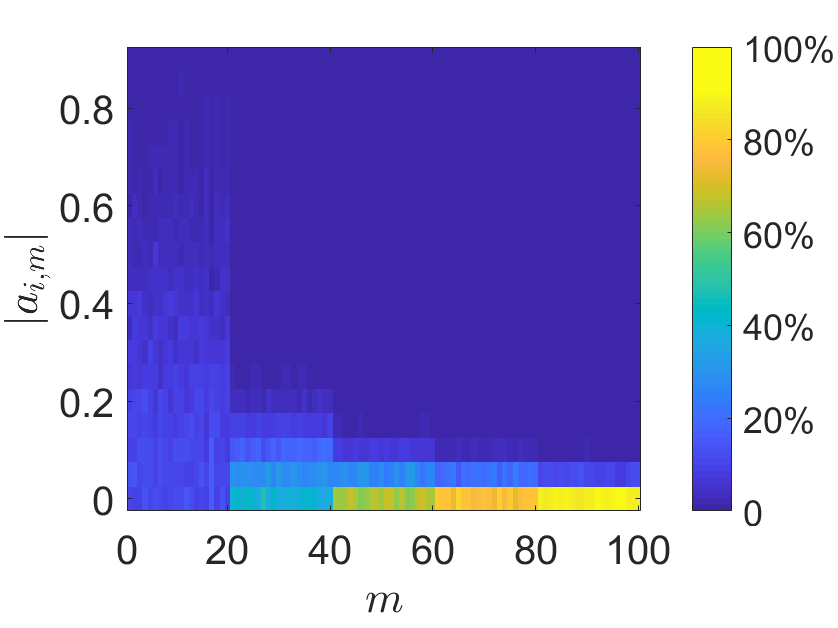}}

\caption{Distributions of ${{a_{i,m}}}$ of different elements in $\boldsymbol{\hat y}$ at the decoder under different parallel-channel scenarios.}
\label{fig:distribution}
\end{figure*}

\begin{figure*}[t!]
\centering 
\setlength{\abovecaptionskip}{0cm}
\setlength{\belowcaptionskip}{-0.5cm}
\subfigbottomskip=7pt 
\subfigcapskip=0pt 
\subfigure[$2 - {\rm{ch}}$.]{
\label{weight2}
\includegraphics[width=.3\linewidth]{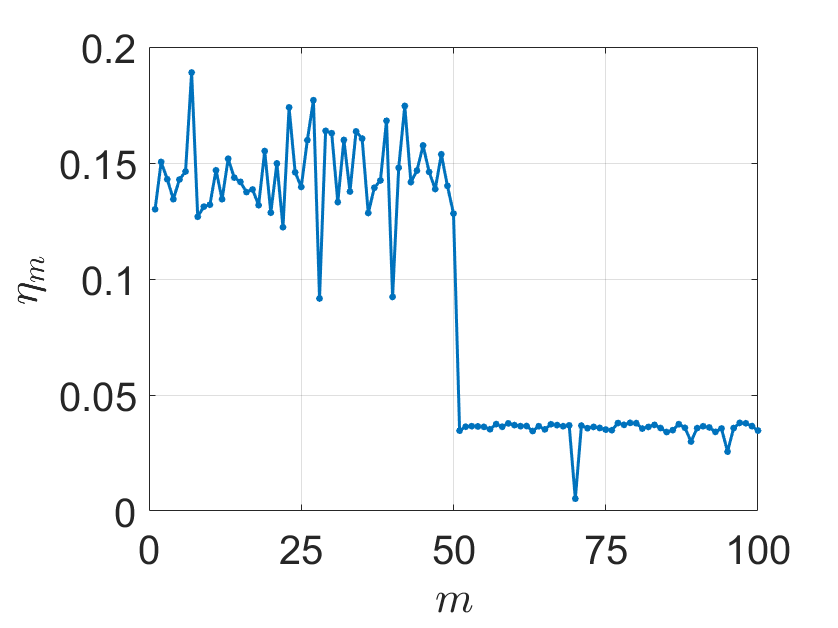}}\hspace{2pt}
\subfigure[$4 - {\rm{ch}}$.]{
\label{weight4}
\includegraphics[width=.3\linewidth]{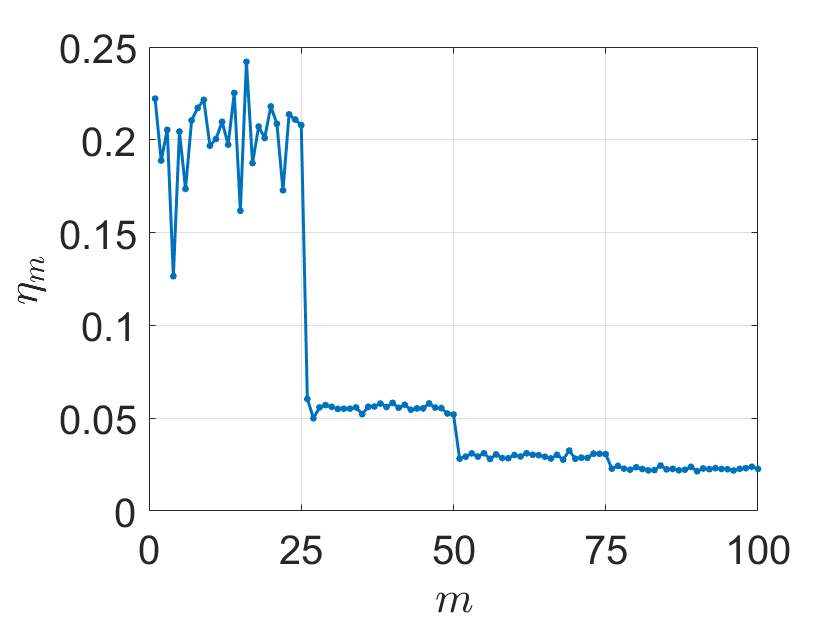}}\hspace{2pt}
\subfigure[$5 - {\rm{ch}}$.]{
\label{weight5}
\includegraphics[width=.3\linewidth]{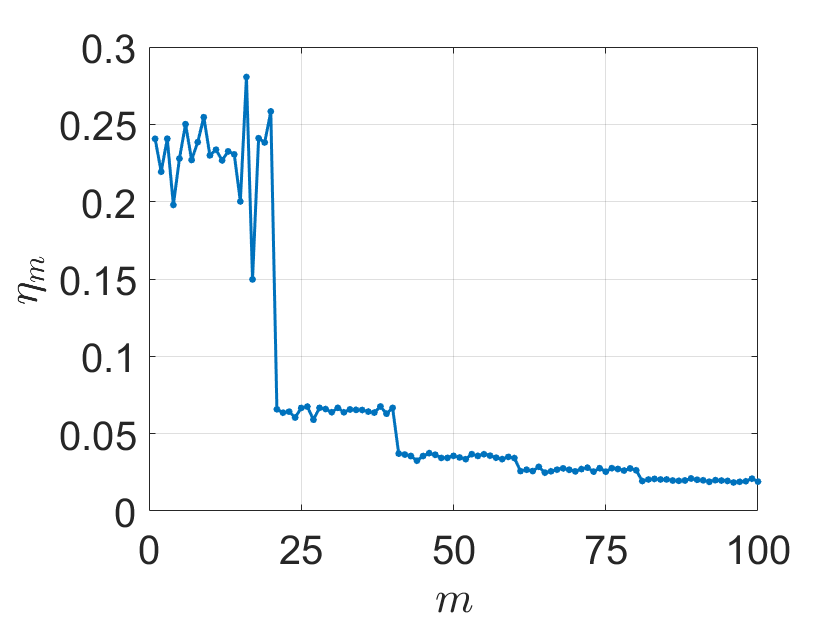}}
\caption{${\eta _m}$ of different elements in $\boldsymbol{\hat y}$ at the decoder under different parallel-channel scenarios.}
\label{fig:weight}
\end{figure*}

Figure \ref{fig:distribution} shows the distributions of ${\left| {{a_{i,m}}} \right|}$, where ${{a_{i,m}}}$ represents the weight of the $i$-th neuron of the first FC layer of the decoder at the $m$-th element. From Fig. \ref{fig:distribution}, we can observe that in the subchannel with small error probability, ${\left| {{a_{i,m}}} \right|}$ randomly appears in the range of $0$ to $0.9$. In contrast, in the subchannel with large error probability, ${\left| {{a_{i,m}}} \right|}$ concentrates around $0$. For example, in Fig. \ref{a_2}, in the subchannels with small error probability ($m \le 50$), ${\left| {{a_{i,m}}} \right|}$ mostly appears in the range of $0$ to $0.2$, and occasionally appears in the range of $0.2$ to $0.5$. In contrast, in the subchannels with large error probability ($m > 50$), ${\left| {{a_{i,m}}} \right|}$ mostly appears in the range of $0$ to $0.05$ and occasionally appears in the range of $0.05$ to $0.1$. This implies that the output of the decoder is mainly calculated according to the elements transmitted through subchannel with small error probability, and the elements transmitted through subchannel with large error probability have little effect on the output of the decoder. This is because AIB-JSCC utilizes the channel condition in the loss function and learns to transmit elements important for reconstruction through the subchannel with small error probability to reduce the loss function. 

Figure \ref{fig:weight} shows the average of ${\left| {{a_{i,m}}} \right|}$, ${\eta _m}$, which is calculated by
\begin{equation}
\label{JSAC-7}
{\eta _m} = \frac{1}{L}\sum\limits_{i = 1}^L {\left| {{a_{i,m}}} \right|},
\end{equation}
where $L$ represents the number of neurons in the first FC layer of the decoder. From Fig. \ref{fig:weight}, we can observe that the elements transmitted through the same subchannel have similar ${\eta _m}$, and the elements transmitted through the subchannel with small error probability have large ${\eta _m}$. For example, in Fig. \ref{weight2}, in the subchannels with small error probability ($m \le 50$), ${\eta _m}$ is in the range of $0.1$ to $0.2$. In contrast, in the subchannels with large error probability ($m > 50$), ${\eta _m}$ is in the range of $0$ to $0.05$, which is much smaller than $0.1$. As analyzed before, this is because AIB-JSCC utilizes the channel condition during training and is able to allocate elements important for reconstruction to subchannel with small error probability.

\begin{figure}[t!]
\centering 
\setlength{\abovecaptionskip}{0cm}
\setlength{\belowcaptionskip}{-0.5cm}
\subfigbottomskip=7pt 
\subfigcapskip=0pt 
\subfigure[Original images.]{
\label{fig:parallel_original}
\includegraphics[width=0.9\linewidth]{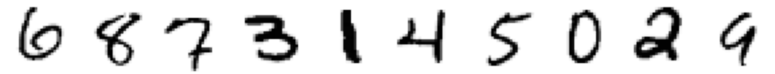}}\\
\subfigure[$2-\rm{ch}$ scenario.]{
\label{fig:parallel2}
\includegraphics[width=0.9\linewidth]{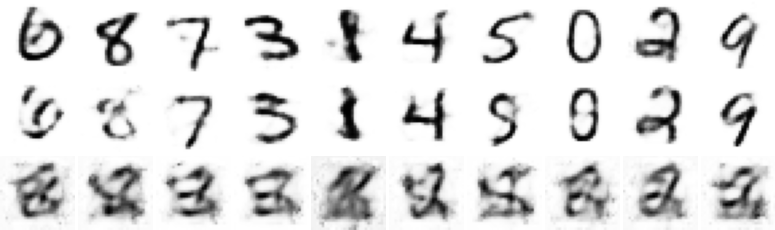}}\\
\subfigure[$5-\rm{ch}$ scenario.]{
\label{fig:parallel5}
\includegraphics[width=0.9\linewidth]{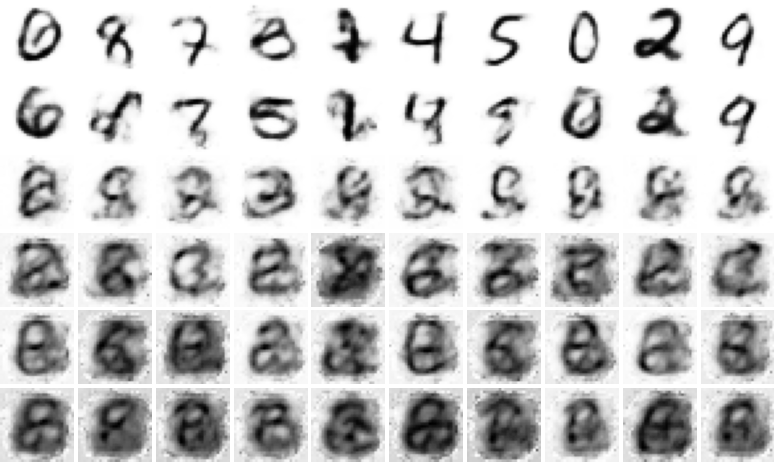}}
\caption{Original image and image recovered in $2-\rm{ch}$ and $5-\rm{ch}$ scenarios. For the $2-\rm{ch}$ and $5-\rm{ch}$ scenarios, the first row is reconstructed from the complete noisy codeword $\boldsymbol{{\boldsymbol{\hat y}}}$. The second and the third rows in $2-\rm{ch}$ scenario is reconstructed from ${{{\boldsymbol{\hat y}}}_{{\rm{ch1}}}}$ and ${{{\boldsymbol{\hat y}}}_{{\rm{ch2}}}}$. The second to the sixth rows in $5-\rm{ch}$ scenario is reconstructed from ${{{\boldsymbol{\hat y}}}_{{\rm{ch1}}}} - {{{\boldsymbol{\hat y}}}_{{\rm{ch5}}}}$.}
\label{fig:parallel}
\end{figure}

Figure \ref{fig:parallel} shows the visual reconstructions of AIB-JSCC recovered from noise codewords received from different subchannels. In particular, when using the noisy codeword received from the $i$-th subchannel, i.e., ${{{\boldsymbol{\hat y}}}_{{\rm{ch}}i}}$, to reconstruct the images, we fix the other elements in $\boldsymbol{\hat y}$ to $0$ and feed the new $\boldsymbol{\hat y}$ into the trained decoder to obtain the reconstructions. From Fig. \ref{fig:parallel}, we can observe that for both $2 - {\rm{ch}}$ and $5 - {\rm{ch}}$ scenarios, the complete noisy codeword $\boldsymbol{\hat y}$ has the best visual performance, and the images recovered from ${{{\boldsymbol{\hat y}}}_{{\rm{ch}}i}}$ that is received from the subchannel with smaller error probability, preserve more semantic information. For instance, the images in the second row in Fig. \ref{fig:parallel2} can be identified easily, while the images in the third row are difficult to recognize. This is because AIB-JSCC utilizes the channel condition in the loss function, and transmitting the elements with more semantic information through the subchannel with smaller error probability is helpful for semantic information preservation and loss reduction. Therefore, AIB-JSCC is able to transmit the elements with more semantic information for reconstruction through the subchannel with small error probability.

\subsection{High-order Scenario}
Table \ref{tab:DMC} shows the MSE of AIB-JSCC and IABF under DMCs with various orders and error probabilities. During training, the orders of the codeword ${\boldsymbol{y}}$ and the noisy codeword ${{\boldsymbol{\hat y}}}$ are set to be identical. The channel transition probability is
\begin{equation}
\label{DSC}
{p_{jl}} = \left\{ {\begin{array}{*{20}{c}}
{1 - \varepsilon }&{j = l}\\
{\frac{\varepsilon }{{Q - 1}}}&{j \ne l}
\end{array}} \right.,
\end{equation}
where $Q$ is the order of ${\boldsymbol{y}}$. From Table \ref{tab:DMC}, we can observe that for identical error probability, when the order of ${\boldsymbol{y}}$ increases, the MSE of AIB-JSCC increases. For instance, when the error probability is $0.1$, the MSE of AIB-JSCC is $57.378$, $58.149$ and $60.712$ when the order of ${\boldsymbol{y}}$  is $3$, $5$, $7$, respectively. This implies that even though using high-order can improve transmission efficiency, it also diminishes performance. Moreover, AIB-JSCC has a lower MSE than IABF under all considered of DMCs and error probabilities. This is because the proposed IB objective preserves semantic information and discards redundant information.

\section{Conclusion}
In this work, we have proposed an AIB-JSCC scheme for image transmission, which can adaptively minimize the transmission rate and distortion at the same time to achieve better reconstruction quality, larger compression ratio, and lower computational complexity than the state-of-the-art approaches. Specifically, we first derived a mathematically tractable form of IB objective for the JSCC system. 
Then, to appropriately balance the reconstruction distortion and the transmission rate, we further proposed an algorithm that can adaptively adjust hyperparameter $\beta$ of the loss function according to the reconstruction error. Experimental results have shown that with fixed length of codewords, AIB-JSCC always achieved smaller reconstruction error than IB-JSCC and IABF over various error probabilities and datasets, which demonstrates the effectiveness of the proposed IB objective and adaptive IB algorithm. In addition, the images recovered by AIB-JSCC had better visual performance and obtained higher accuracy on downstream classification task than IABF. For a given reconstruction error, AIB-JSCC always permitted larger compression ratio than SSCC and IABF. In particular, AIB-JSCC only needed around $4\%$ and $80\%$ as many elements compared with SSCC and IABF. Moreover, AIB-JSCC also had lower computational complexity and was more 
robust than IABF. In the parallel-channel scenarios, AIB-JSCC was able to transmit elements important for reconstruction in the subchannel with small error probability. The overall results showed that the proposed schemes can significantly reduce the transmission rate, and improve the reconstruction quality and downstream task accuracy with lower computational complexity.

\begin{table}[t!]
\renewcommand\arraystretch{1.4}
\caption{\label{tab:DMC}MSE of recovered images when the channel is DMC}
\begin{center}
\begin{tabular}{|c|c|c|c|c|c|}
\hline
\multirow{3}*{\textbf{Order}} & \multirow{3}*{\textbf{Methods}} & \multicolumn{4}{c|}{\textbf{MSE under different} } \\
~ & ~ & \multicolumn{4}{c|}{\textbf{ error probabilities}} \\
\cline{3-6}
~ & ~ & \textbf{0.1} & \textbf{0.2} & \textbf{0.3} & \textbf{0.4}\\
\hline
\multirow{2}*{3} & IABF & 58.478
 & 65.227 & 72.543 & 81.664  \\
\cline{2-6}
~ & AIB-JSCC & \textbf{57.378} & \textbf{65.054} & \textbf{72.231} & \textbf{79.797} \\
\hline
\multirow{2}*{5} & IABF & 59.028 &	66.815 & 74.381 & 81.92 \\
\cline{2-6}
~ & AIB-JSCC & \textbf{58.149} & \textbf{65.363} & \textbf{73.015} & \textbf{80.859} \\
\hline
\multirow{2}*{7} & IABF & 61.901 &	70.097 & 76.169 & 85.379 \\
\cline{2-6}
~ & AIB-JSCC & \textbf{60.712} & \textbf{68.463} & \textbf{75.777} & \textbf{85.175} \\
\hline
\end{tabular}
\end{center}
\end{table}
\appendix[PROOF OF LEMMA \ref{le:1}]
Let ${\rm{I}}{{\rm{B}}_\beta }\left( {{\boldsymbol{x,y,\hat y}}} \right) = I\left( {{\boldsymbol{x}};{\boldsymbol{\hat y}}} \right) - \beta I\left( {{\boldsymbol{x}};{\boldsymbol{y}}} \right)$. We need to guarantee that ${\rm{I}}{{\rm{B}}_\beta }\left( {\boldsymbol{x},\boldsymbol{y},\boldsymbol{\hat y}} \right)$ is not maximal when $\boldsymbol{x}$ and $\boldsymbol{y}$ are independent, i.e., $p\left( {\boldsymbol{y}|\boldsymbol{x}} \right) = p\left( \boldsymbol{y} \right)$ or $p\left( {\boldsymbol{x}|\boldsymbol{y}} \right) = p\left( \boldsymbol{x} \right)$ is not optimal for maximizing ${\rm{I}}{{\rm{B}}_\beta }\left( {{\boldsymbol{x}},{\boldsymbol{y}},{\boldsymbol{\hat y}}} \right)$. Since there is a Markov chain relationship $\boldsymbol{x} \to \boldsymbol{y} \to {\boldsymbol{\hat y}}$, we have $I\left( {\boldsymbol{x};\boldsymbol{y}} \right) \ge I\left( {\boldsymbol{x};{\boldsymbol{\hat y}}} \right)$. When $\boldsymbol{x}$ and $\boldsymbol{y}$ are independent, $I\left( {\boldsymbol{x};\boldsymbol{y}} \right) = I\left( {\boldsymbol{x};{\boldsymbol{\hat y}}} \right) = 0$ and ${\rm{I}}{{\rm{B}}_\beta }\left( {\boldsymbol{x},\boldsymbol{y},{\boldsymbol{\hat y}}} \right){|_{p\left( {\boldsymbol{y}|\boldsymbol{x}} \right) = p\left( \boldsymbol{y} \right)}} = 0$ for any $\beta$. Therefore, if ${\rm{I}}{{\rm{B}}_{{\beta _1}}}\left( {{\boldsymbol{x}},{\boldsymbol{y}},{\boldsymbol{\hat y}}} \right)$ is not maximal when ${p\left( {{\boldsymbol{y}}|{\boldsymbol{x}}} \right) = p\left( {\boldsymbol{y}} \right)}$, there must exist $\left( {\boldsymbol{x},\boldsymbol{y},{\boldsymbol{\hat y}}} \right)$ given by ${p_1}\left( {\boldsymbol{y}|\boldsymbol{x}} \right)$ such that 
\begin{equation}
\label{1_6_1}
{\rm{I}}{{\rm{B}}_{{\beta _1}}}\left( {\boldsymbol{x},\boldsymbol{y},{\boldsymbol{\hat y}}} \right){|_{p\left( {\boldsymbol{y}|\boldsymbol{x}} \right) = {p_1}\left( {\boldsymbol{y}|\boldsymbol{x}} \right)}} > {\rm{I}}{{\rm{B}}_\beta }\left( {\boldsymbol{x},\boldsymbol{y},{\boldsymbol{\hat y}}} \right){|_{p\left( {\boldsymbol{y}|\boldsymbol{x}} \right) = p\left( \boldsymbol{y} \right)}} = 0.
\end{equation}
If ${\rm{I}}{{\rm{B}}_{{\beta _1}}}\left( {\boldsymbol{x},\boldsymbol{y},{\boldsymbol{\hat y}}} \right)$ is not optimal when ${p\left( {{\boldsymbol{y}}|{\boldsymbol{x}}} \right) = p\left( {\boldsymbol{y}} \right)}$, we can rewrite (\ref{1_6_1}) as
\begin{equation}
\label{1_6_2}
I\left( {\boldsymbol{x};{\boldsymbol{\hat y}}} \right) - {\beta _1}I\left( {\boldsymbol{x};\boldsymbol{y}} \right) > 0,\exists \boldsymbol{x},\boldsymbol{y},{\boldsymbol{\hat y}}.
\end{equation}
According to (\ref{1_6_2}), we have
\begin{equation}
\label{1_6_3}
\beta _1  < {\beta _0} = \mathop {\sup }\limits_{\boldsymbol{x} \to \boldsymbol{y} \to {\boldsymbol{\hat y}}} \frac{{I\left( {\boldsymbol{x};{\boldsymbol{\hat y}}} \right)}}{{I\left( {\boldsymbol{x};\boldsymbol{y}} \right)}}.
\end{equation}
This completes the proof.$\hfill\square$

\bibliographystyle{IEEEtran}
\bibliography{IEEEabrv}

\vspace{-3ex}
\section{Biography Section}
\begin{IEEEbiography}[{\includegraphics[width=1in,height=1.25in,clip,keepaspectratio]{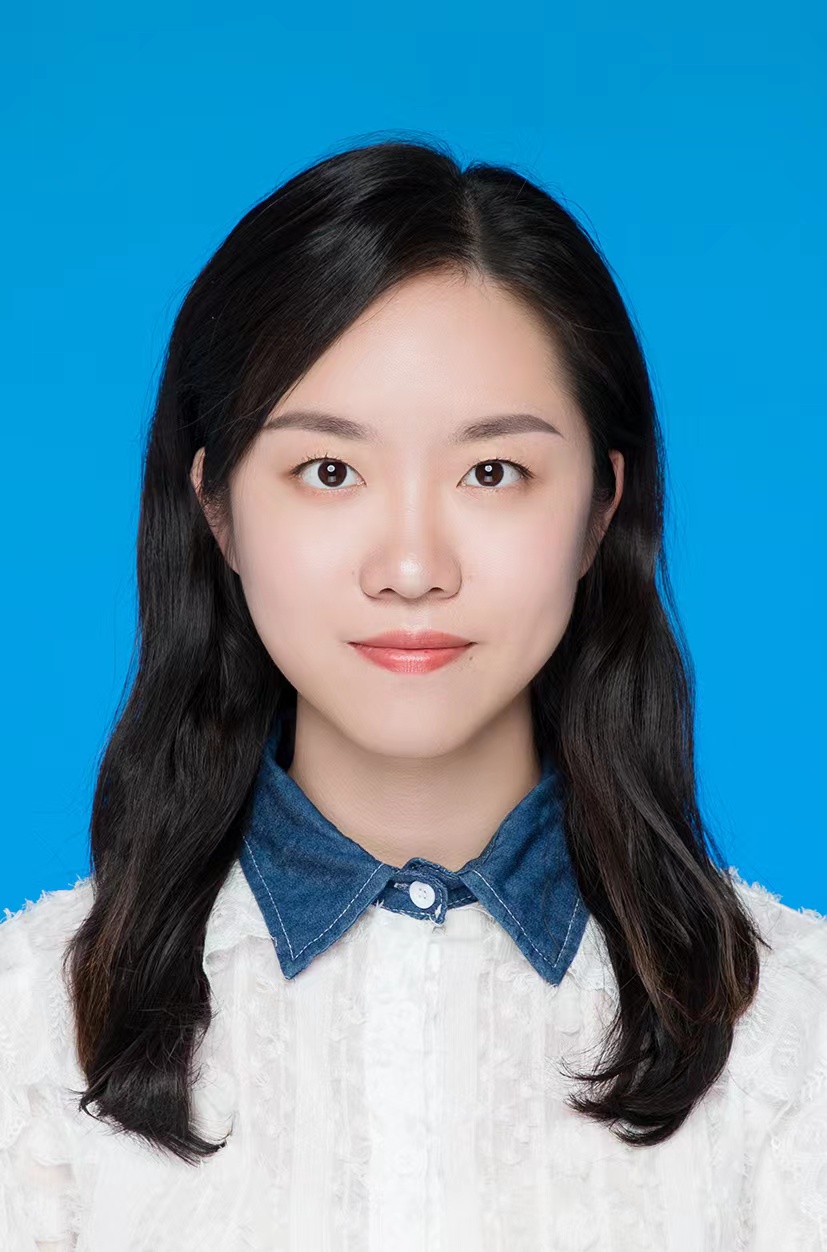}}]{Lunan Sun} received the B.S. degree from Beijing Jiaotong University, Beijing, China, in 2018. She is currently pursuing the Ph.D. degree with the Beijing Key Laboratory of Network System Architecture and Convergence, Beijing University of Posts and Telecommunications, Beijing, China. Her current research interests include semantic communications, image transmission and deep learning.
\end{IEEEbiography}
\vspace{-3ex}
\begin{IEEEbiography}[{\includegraphics[width=1in,height=1.25in,clip,keepaspectratio]{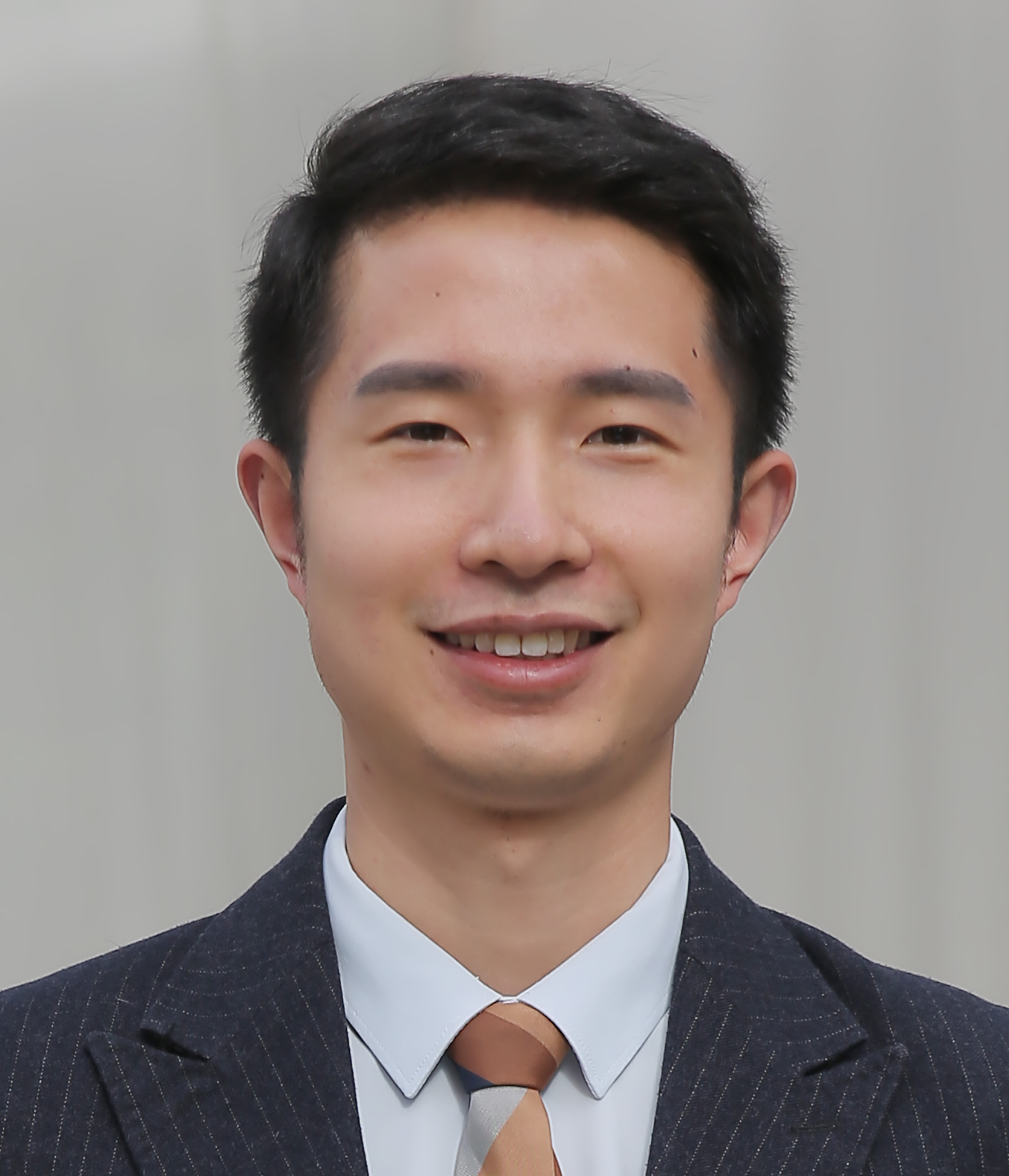}}]{Yang Yang} (Member, IEEE) is currently an Associate Professor with the School of Information and Communication Engineering, Beijing University of Posts and Telecommunications (BUPT), Beijing, China. He received the B.S. degree in information engineering from School of Communication, Xidian University, Xian, China, in June 2013, and the Ph.D degree in Information and Communication Engineering from School of Information and Communication Engineering, Beijing University of Posts and Telecommunications (BUPT), Beijing, China. His research interests include semantic communications, visible light communication and localization. He served as a workshop co-chair/TPC member for a series of IEEE conferences including Globecom, ICC and WCNC. He was a recipient of the IEEE Wireless Communications and Networking Conference (WCNC) 2021 Best Paper Award.
\end{IEEEbiography}
\vspace{-3ex}
\begin{IEEEbiography}[{\includegraphics[width=1in,height=1.25in,clip,keepaspectratio]{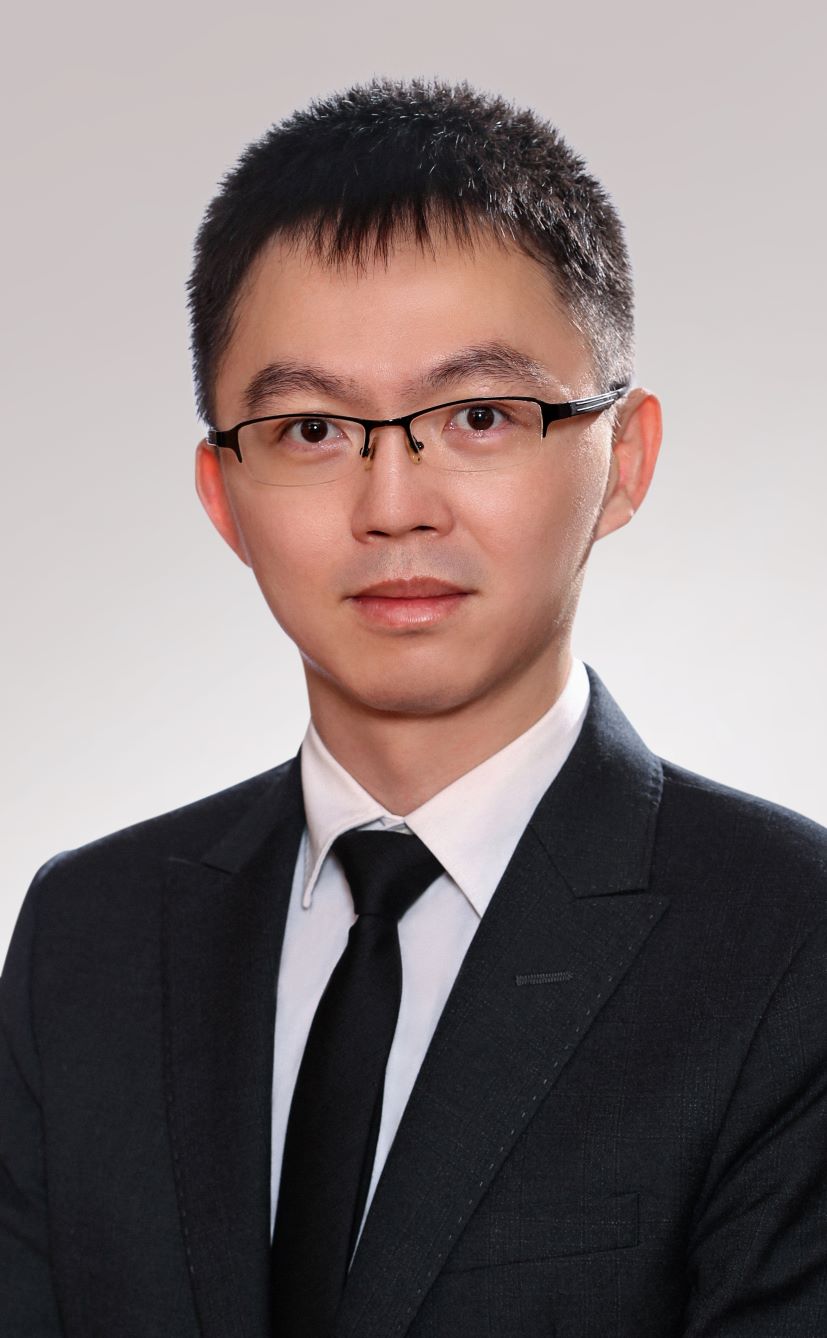}}]{Mingzhe Chen} (Member, IEEE) is currently an Assistant Professor with the Department of Electrical and Computer Engineering and Institute of Data Science and Computing at University of Miami. His research interests include federated learning, reinforcement learning, virtual reality, unmanned aerial vehicles, and Internet of Things. He has received from the IEEE Communication Society four journal paper awards including the IEEE Marconi Prize Paper Award in Wireless Communications in 2023, the Young Author Best Paper Award in 2021 and 2023, and the Fred W. Ellersick Prize Award in 2022, and three conference best paper awards at IEEE ICC in 2020, IEEE GLOBECOM in 2020, and IEEE WCNC in 2021. He currently serves as an Associate Editor of IEEE Transactions on Mobile Computing, IEEE Wireless Communications Letters, IEEE Transactions on Green Communications and Networking, and IEEE Transactions on Machine Learning in Communications and Networking.
\end{IEEEbiography}
\vspace{-3ex}
\begin{IEEEbiography}[{\includegraphics[width=1in,height=1.25in,clip,keepaspectratio]{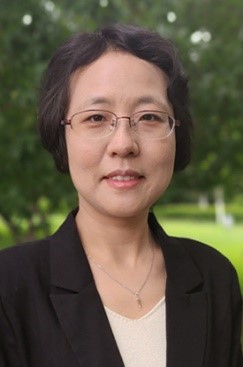}}]{Caili Guo} (Senior Member, IEEE) received the Ph.D. degree in Communication and Information Systems from Beijing University of Posts and Telecommunication (BUPT) in 2008. She is currently a Professor in the School of Information and Communication Engineering at BUPT. Her general research interests include machine learning and statistical signal processing for wireless communications, with current emphasis on semantic communications, deep learning, and intelligence-enabled edge computing for vehicle communications.

In the related areas, she has published over 200 papers and holds over 30 granted patents. She won Diamond Best Paper Award of IEEE ICME 2018 and Best Paper Award of IEEE WCNC 2021.
\end{IEEEbiography}
\vspace{-3ex}
\begin{IEEEbiography}[{\includegraphics[width=1in,height=1.25in,clip,keepaspectratio]{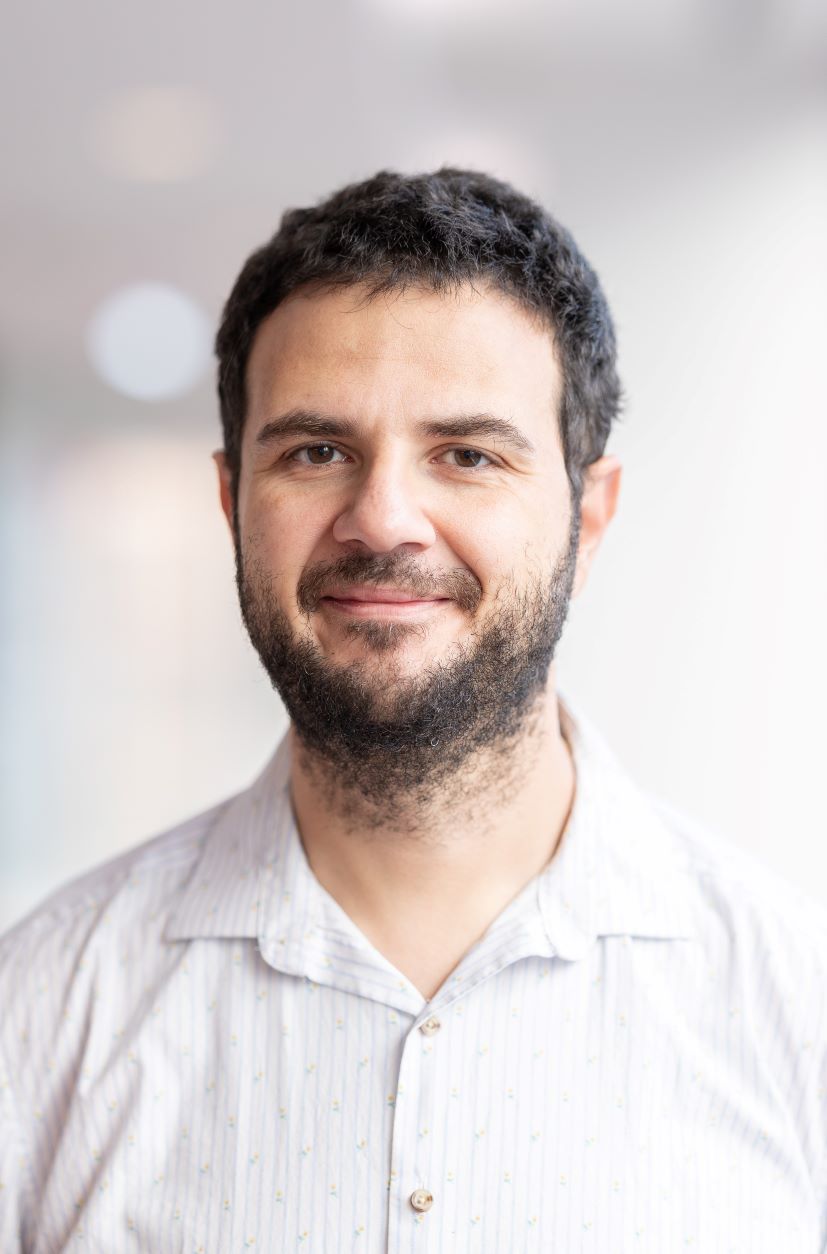}}]{Walid Saad}  (Fellow, IEEE) received his Ph.D degree from the University of Oslo, Norway in 2010. He is currently a Professor at the Department of Electrical and Computer Engineering at Virginia Tech, where he leads the Network sciEnce, Wireless, and Security (NEWS) laboratory. He is also the Next-G Wireless Faculty Lead at Virginia Tech's Innovation Campus. His research interests include wireless networks (5G/6G/beyond), machine learning, game theory, security, UAVs, semantic communications, cyber-physical systems, and network science. Dr. Saad is a Fellow of the IEEE. He is also the recipient of the NSF CAREER award in 2013, the AFOSR summer faculty fellowship in 2014, and the Young Investigator Award from the Office of Naval Research (ONR) in 2015. He was the (co-)author of eleven conference best paper awards at IEEE WiOpt in 2009, ICIMP in 2010, IEEE WCNC in 2012, IEEE PIMRC in 2015, IEEE SmartGridComm in 2015, EuCNC in 2017, IEEE GLOBECOM (2018 and 2020), IFIP NTMS in 2019, IEEE ICC (2020 and 2022). He is the recipient of the 2015 and 2022 Fred W. Ellersick Prize from the IEEE Communications Society,  and of the IEEE Communications Society Marconi Prize Award in 2023. He was also a co-author of the papers that received the IEEE Communications Society Young Author Best Paper award in 2019, 2021, and 2023. Other recognitions include the 2017 IEEE ComSoc Best Young Professional in Academia award, the 2018 IEEE ComSoc Radio Communications Committee Early Achievement Award, and the 2019 IEEE ComSoc Communication Theory Technical Committee Early Achievement Award. From 2015-2017, Dr. Saad was named the Stephen O. Lane Junior Faculty Fellow at Virginia Tech and, in 2017, he was named College of Engineering Faculty Fellow. He received the Dean's award for Research Excellence from Virginia Tech in 2019. He was also an IEEE Distinguished Lecturer in 2019-2020.  He has been annually listed in the Clarivate Web of Science Highly Cited Researcher List since 2019. He currently serves as an Area Editor for the IEEE Transactions on Network Science and Engineering and the IEEE Transactions on Communications. He is the Editor-in-Chief for the IEEE Transactions on Machine Learning in  Communications and Networking.
\end{IEEEbiography}
\vspace{-3ex}
\begin{IEEEbiography}[{\includegraphics[width=1in,height=1.25in,clip,keepaspectratio]{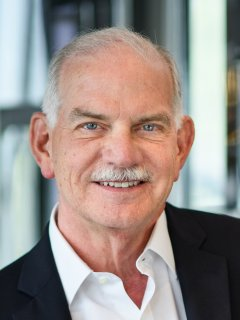}}]{H. Vincent Poor}  (Life Fellow, IEEE) received the Ph.D. degree in EECS from Princeton University in 1977.  From 1977 until 1990, he was on the faculty of the University of Illinois at Urbana-Champaign. Since 1990 he has been on the faculty at Princeton, where he is currently the Michael Henry Strater University Professor. During 2006 to 2016, he served as the dean of Princeton’s School of Engineering and Applied Science. He has also held visiting appointments at several other universities, including most recently at Berkeley and Cambridge. His research interests are in the areas of information theory, machine learning and network science, and their applications in wireless networks, energy systems and related fields. Among his publications in these areas is the recent book Machine Learning and Wireless Communications.  (Cambridge University Press, 2022). Dr. Poor is a member of the National Academy of Engineering and the National Academy of Sciences and is a foreign member of the Chinese Academy of Sciences, the Royal Society, and other national and international academies. He received the IEEE Alexander Graham Bell Medal in 2017.
\end{IEEEbiography}

\vspace{11pt}

\vfill

\end{document}